\documentclass[12pt,preprint]{aastex}
\usepackage{graphicx}

\begin{document}
\shortauthors{Pandian et al.}

\title{The Arecibo Methanol Maser Galactic Plane Survey--IV: Accurate Astrometry and Source Morphologies}

\author{J. D. Pandian\altaffilmark{1}, E. Momjian\altaffilmark{2}, Y. Xu\altaffilmark{3}, K. M. Menten\altaffilmark{4} and P. F. Goldsmith\altaffilmark{5}}
\altaffiltext{1}{Institute for Astronomy, University of Hawaii, 2680 Woodlawn Dr., Honolulu, HI 96822; jpandian@ifa.hawaii.edu}
\altaffiltext{2}{National Radio Astronomy Observatory, P.O. Box O, Socorro, NM 87801}
\altaffiltext{3}{Purple Mountain Observatory, Chinese Academy of Sciences, Nanjing 210008, China}
\altaffiltext{4}{Max-Planck-Institut f\"{u}r Radioastronomie, Auf dem H\"{u}gel 69, 53121 Bonn, Germany}
\altaffiltext{5}{Jet Propulsion Laboratory, California Institute of Technology, Pasadena, CA 91109}

\begin{abstract}
We present accurate absolute astrometry of 6.7~GHz methanol masers detected in the Arecibo Methanol Maser Galactic Plane Survey using MERLIN and the Expanded Very Large Array (EVLA). We estimate the absolute astrometry to be accurate to better than 15 and 80 milliarcseconds for the MERLIN and EVLA observations respectively. We also derive the morphologies of the maser emission distributions for sources stronger than $\sim 1$~Jy. The median spatial extent along the major axis of the regions showing maser emission is $\sim 775$~AU. We find a majority of methanol maser morphologies to be complex with some sources previously determined to have regular morphologies in fact being embedded within larger structures. This suggests that some maser spots do not have a compact core, which leads them being resolved in high angular resolution observations. This also casts doubt on interpretations of the origin of methanol maser emission solely based on source morphologies. We also investigate the association of methanol masers with mid-infrared emission and find very close correspondence between methanol masers and 24~$\mu$m point sources. This adds further credence to theoretical models that predict methanol masers to be pumped by warm dust emission and firmly reinforces the finding that Class~II methanol masers are unambiguous tracers of embedded high-mass protostars.
\end{abstract}


\section{Introduction}

The 6668.519~MHz $J_K = 5_0-6_1$~A$^+$ transition is the strongest and most widespread Class II methanol maser line \citep{ment91}. There is strong evidence that 6.7~GHz methanol masers trace very early evolutionary phases of high-mass star formation. For example, multi-wavelength studies of 6.7~GHz methanol masers have found their environment to be characteristic of massive protoclusters \citep{mini05}. Moreover, recent work modeling the spectral energy distributions of the masers' host sources have found them to be consistent with that of massive young stellar objects (MYSOs) undergoing rapid accretion \citep{pand10}.

However, the question where the maser emission originates relative to the MYSO is still the subject of considerable debate. Early studies (e.g. \citealt{norr93,phil98}) found a number of methanol masers to have linear or arched morphologies with monotonic velocity gradients across the spatial features. These were interpreted to be indicative of the masers originating in an edge-on circumstellar disk. However, the minimum mass for the central object derived from the disk hypothesis was found to be unrealistically high in some sources \citep{wals98}. In contrast, assuming that the linear maser distributions trace the full extent of such a disk, the enclosed mass was found to be much less than 1 M$_\sun$ in a number of sources \citep{mini00}. This led to suggestions that the masers only trace partial disks around massive stars, or are associated with outflows or shock fronts. \citet{dods04} suggested that methanol masers originated behind a planar shock propagating through a star forming core in order to explain the observation of velocity gradients within individual spot clusters that were perpendicular to the main large scale velocity gradient across the linear morphology. 

High spatial resolution mid-infrared studies suggest that many methanol masers, including some with linear morphologies, are associated with outflows rather than being located in disks. While G35.20--0.74 is the best example of masers occurring along the walls of an outflow cavity \citep{debuiz06}, several other masers  were found to have their linear morphologies aligned parallel to outflows \citep{debuiz03}. However, there are also sources such as NGC 7538 IRS1 \citep{pest09} where there is strong evidence that masers originate in a disk. Furthermore, in the recent study of the proper motions of 12.2~GHz methanol masers (which are the second strongest Class II methanol masers) in W3(OH), \citet{mosc10} find that the linear distribution of the masers can be well fit with a flat rotating disk that is seen almost edge-on. Since there is very good spatial correlation between the locations of 6.7 and 12.2~GHz methanol masers, this can be taken as evidence for a disk origin of some 6.7~GHz methanol masers.

Recent high resolution imaging of 6.7~GHz methanol masers with the European VLBI network (EVN) revealed a number of masers to have ring shaped morphologies \citep{bart09}. The ring morphologies were modeled as an inclined disk or torus around a MYSO having expanding or infalling kinematics. In addition, a number of masers were found to be associated with sources that had excess emission measured in the 4.5 $\mu$m IRAC band with the Spitzer space telescope (see \citealt{cyga09} for a detailed discussion of these sources called Extended Green Objects or EGOs). Taken together, this was interpreted as evidence for the origin of methanol maser emission in the interface region between a disk/torus and an outflow.

One of the potential problems in very high angular resolution studies is that a significant fraction of the maser flux is resolved out. For example, \citet{mini02} found that a number of maser spots had a compact core and a more diffuse halo, while some spectral features had no detectable compact core. Hence, lower angular resolution observations are required to determine the full extent and morphology of maser emission. It is also of interest to carry out this work towards a large and homogeneous sample from a blind survey.

The most sensitive blind survey to date for 6.7~GHz methanol masers is the Arecibo Methanol maser Galactic Plane Survey (AMGPS; \citealt{pand07a}, hereafter Paper I). AMGPS detected 86 6.7~GHz methanol masers, of which 48 were new detections. Taking into account the pointing accuracy of the Arecibo telescope, the uncertainty in the position of the AMGPS masers is $\sim 18\arcsec$ at the 95\% confidence level \citep[Paper II hereafter]{pand07b}. Paper II also presented an analysis of the properties of the masers including association with mid-infrared sources, a number of methanol masers were found to not have mid-infrared counterparts in the Galactic plane survey using the Midcourse Source Experiment (MSX; \citealt{pric01}). \citet[Paper III hereafter]{pand09} presented the systemic velocities and distances to the sources (all the distances used in this paper are taken from Paper III) leading to the determination of the luminosity function of 6.7~GHz methanol masers. In this paper, we present accurate astrometry for the AMGPS sources using MERLIN\footnote{Based on observations made with MERLIN, a National Facility operated by the University of Manchester at Jodrell Bank Observatory on behalf of STFC.} and the Expanded Very Large Array\footnote{The National Radio Astronomy Observatory is a facility of the National Science Foundation operated under cooperative agreement by Associated Universities, Inc.} (EVLA). We also derive morphologies of the maser emission for the strong sources (peak flux densities $\gtrsim 1$~Jy), and examine the association of methanol masers with mid-infrared sources found in recent sensitive surveys using the Spitzer space telescope.

\section{Observations and data reduction}
\subsection{MERLIN observations}\label{merlinobs}
The MERLIN observations of the AMGPS masers (W51 main and G41.87--0.10 excluded) were carried out between 2007 February and June using six antennas except for observations between March 5 and March 11 which employed only five antennas. Two correlator configurations were used for the observations. In the wide-band mode, two polarizations were observed with a total bandwidth of 16 MHz split into 32 channels with a channel width of 0.5 MHz. In the narrow-band mode, two polarizations were observed with a bandwidth of 1 MHz split into 256 channels yielding a channel spacing of 3.9~kHz. The primary and bandpass calibrator, 3C84 was observed in both wide and narrow-band modes. The phase calibrators (see Table~\ref{merlincalib}) were only observed in wide-band mode to achieve adequate signal to noise ratio, while the target sources were only observed in narrow-band mode. Phase referencing was used with each scan on a target source being preceded and followed by a scan on a phase calibrator. To achieve good uv-coverage, the target observing sequence was cycled through a group of sources with similar systemic velocities. No Doppler tracking was used, and each group of targets was observed at a fixed frequency. This observing sequence was repeated several times over the course of each observing run which typically lasted for $\sim 10$ hours. The final integration time was roughly 45 minutes per source, with the resulting $1\sigma$ channel noise being $\sim 45$ mJy beam$^{-1}$. At the adopted rest frequency of 6668.519 MHz for the maser transition, the velocity resolution in the narrow-band mode is 0.18~km~s$^{-1}$.

The data were initially reduced using local MERLIN software \citep{diam03}. After initial editing and flux calibration on 3C84, the data were converted into FITS format with subsequent processing being carried out using the Astronomical Image Processing System (AIPS) package of NRAO. The flux calibration was carried out assuming the flux density of 3C84 to be 14.5~Jy in 2007 February, 15.0~Jy in 2007 March, 15.5~Jy in 2007 April, 16.0~Jy in 2007 May, and 16.5~Jy in 2007 June. The phases of 3C84 were calibrated in both the wide-band and narrow-band data, with the derived phase offset between the two data sets being used to transfer the calibration from the phase reference source (which is only observed in the wide-band mode) to the target sources (which are only observed in the narrow-band mode). After calibration, the task ``CVEL'' was used to shift the spectra to the systemic velocity with respect to the local standard of rest (LSR) and correct for the effects of Earth's motion. The target spectra were then inspected followed by imaging the brightest channel to obtain the target astrometry. The typical beamsize was $60 \times 35$ mas with a position angle of 20\degr.

As indicated in Sect. 2.3.1 of \citet{bart09}, the astrometric accuracy is limited by four factors -- the accuracy of the phase calibrator position, accuracy in the positions of the individual antennas, accuracy in the transfer of phase solutions from the phase calibrator and targets (on account of the atmosphere), and the precision with which the emission location can be determined given the beamsize. The phase calibrator positions indicated in Table~\ref{merlincalib} have position uncertainties between 0.4 and 4.4 milliarcseconds (mas). The antenna positions have uncertainties of 1--2 cm, which results in a position uncertainty of $\sim 10$ mas. To determine the uncertainties from the phase transfer from the phase calibrator to the targets, we inspected the phase change over a timescale corresponding to the separation between the calibrator and target (e.g. an angular separation of 3\degr~corresponds to a 12 min timescale). We estimate positional uncertainties to be less than 10 mas, with median values of 6 mas. The last uncertainty from formal fitting errors is usually $\ll 1$ mas, and can usually be ignored. Combining the various uncertainties, we estimate the absolute astrometric accuracy to be better than 15 mas.

\subsection{EVLA observations}\label{evlaobs}
Among the 81 AMGPS masers targeted in the MERLIN observations, 30 sources were either not detected, or had bad data. After verification of the peak flux densities of these sources with the 100 m Effelsberg telescope to check for variability, we observed 27 sources in 24 fields using the EVLA. The observations were carried out on 2008 October 16 and 18 in the A configuration. A single polarization (RR) was recorded using the old VLA correlator with a bandwidth of 1.56~MHz and 512 spectral channels. Since the lower 0.5~MHz of the bandpass was unusable due to aliasing, this setup was required to achieve the highest possible spectral resolution along with the required velocity coverage. The resulting channel spacing was 3~kHz giving a velocity resolution of 0.14~km~s$^{-1}$ at the rest frequency of the maser transition.

Each target source was observed in a single 10 minute or two 15 minute scans depending on its single dish flux density, with the resulting $1\sigma$ channel noise ranging from 20 to 35 mJy beam$^{-1}$. The sky frequencies were computed for each individual scan and placed within the upper two-thirds of the 1.56 MHz bandwidth to ensure that no maser component would fall in the lower 0.5 MHz of the bandpass. The source J1331+305 (3C286) was used to set the absolute flux density scale, and the sources J1751+096 and J1925+211 as bandpass calibrators. The phase calibrators, used to calibrate the complex gains were J1851+005, J1856+061 and J1922+155 (Table~\ref{evlacalib}), with each calibrator being used for a subset of the target sources closest in angular distance.

The editing, calibration, and processing of the data were carried out using AIPS. After the transfer of the complex gain calibration solutions to the target sources, the strongest emission channel in each source was imaged and its position measured followed by self-calibration (when sufficient signal to noise was available) in both phase and amplitude in an iterative cycle. In several cases, multi-field imaging was necessary to account for strong 6.7~GHz methanol maser emission from other sources that fell within the primary beam of the 25 m EVLA antennas. In some weak sources, multiple channels were smoothed in order to detect the source.

The phase calibrators used in the EVLA observations have a position accuracy better than 3 mas (Table~\ref{evlacalib}). The EVLA antenna positions are known to an accuracy of $\sim 1$ cm implying a position uncertainty of 55 mas. The uncertainty in the astrometry resulting from the transfer of phase solutions from the phase calibrator to the target (obtained from examining the raw phases of the phase calibrator) is estimated to be less than 30 mas. In addition, some sources are only detected with moderate to poor signal to noise ratio, and consequently the formal fitting errors in determining the position of peak emission are as high as $\sim 45$ mas for these sources. Thus, we estimate the absolute astrometry to be accurate to between 65 and 80 mas.

\subsection{24~$\mu$m counterparts}\label{mipscounterparts}
To examine the relation between 6.7~GHz methanol masers and mid-infrared emission, we looked for 24~$\mu$m counterparts in the Spitzer MIPSGAL survey \citep{care09}. Since the 24~$\mu$m point source catalog has not been released, we used the mosaiced images corrected for artifacts for this purpose. The strong interstellar dust emission in the Galactic plane leads to complex backgrounds in the images, which can significantly affect the measurement of the positions of point sources. Hence, we used a $11 \times 11$ pixel median filter to subtract the background, as suggested in the MIPSGAL data reduction cookbook\footnote{http://ssc.spitzer.caltech.edu/dataanalysistools/cookbook/30/\#\_Toc272753806}. The use of such aggressive filtering reduces the source flux to some extent. While this can be accounted for, we are primarily interested in the positions of the point sources in this work. Hence, we do not derive flux densities of the 24~$\mu$m counterparts.

An additional complication arises from many regions being extremely bright and saturated in the MIPSGAL data. While some regions are completely blanked and other sources have the entire core of the point spread function (PSF) blanked, there are some sources for which only the central few pixels are blanked. Although the latter are rejected in the point source detection algorithm of MOPEX, there is sufficient information to measure the positions of these sources. Hence, we measured the positions of all 24~$\mu$m counterparts by Gaussian fitting of the PSF cores using the routine ``JMFIT'' in AIPS. Although the full width at half maximum (FWHM) of the MIPS PSF at 24~$\mu$m is 5.9\arcsec, the high signal to noise ratio of the counterparts implies that the formal uncertainties in the Gaussian fits are small, and that systematic errors dominate the uncertainties in the positions derived for the MIPSGAL counterparts.

\section{Results}

We were able to measure the absolute positions of 57 sources using MERLIN (Table~\ref{merlinpos}; positions of masers in W51 main were obtained using the data of \citealt{xu09}) and an additional 25 sources using the EVLA (Table~\ref{evlapos}). We also identified two masers in the MERLIN data that were not in the original AMGPS catalog (one of these sources was detected by \citealt{cyga09}). Both sources are located close to other sources, leading them to be either missed or unresolved in the original survey. Thus, this work has determined the absolute positions to 82 out of 88 AMGPS sources.

Point source counterparts at 24~$\mu$m were found for 60 out of 82 sources. 18 sources were either fully saturated or in the middle of a blanked region. Another two sources was found to be in the middle of extended emission, while no counterparts were found for two sources. The positions of the 24~$\mu$m point source counterparts are tabulated in Table~\ref{mipspos}, and comments about selected sources are included in Sect. \ref{indisources}.

\subsection{Relative astrometry and spot morphology}\label{relast}

Although the absolute astrometric accuracy is limited by the factors described in Sect. \ref{merlinobs} and \ref{evlaobs}, the relative astrometry between the emission distributions in different velocity channels can be determined to much higher precision. To this end, we imaged each source over the full velocity range of its emission, and measured the positions of the maser spots relative to the channel with peak emission. To improve the dynamic range in the data cubes, we self-calibrated data of targets that were sufficiently strong (peak flux density $\gtrsim 1$~Jy). The measurement of spot positions in the data cube was automated using the AIPS task ``SAD''. To obtain reliable fits, we imposed a 6$\sigma$ threshold on both the peak and integrated flux density to qualify as a maser spot.

Figs. \ref{merlinspotmaps} and \ref{evlaspotmaps} show the results of this work for the MERLIN and EVLA data respectively. For each source, the top panel shows the integrated spectrum, while the bottom panel shows the morphology of the maser spot emission as a function of both angular and physical distance. The coordinates listed in Tables \ref{merlinpos} and \ref{evlapos} serve as reference coordinates for the spot maps. The LSR velocities of the maser spots are indicated next to the spots themselves. When a group of spots form a morphology with a monotonic velocity gradient, only the initial and final velocities are indicated at the ends of the feature. In cases where there is a compact cluster of spots, or where the spots form a morphology without a monotonic velocity gradient, the velocity range of the cluster is indicated next to the feature. Some sources observed with MERLIN are detected in three spots or fewer, from which no morphological information can be deduced. For these cases, only the integrated spectra are shown in Fig.~\ref{merlinspectra}. The EVLA data have much poorer resolution compared to the MERLIN data, and thus spot morphologies can be deduced only for the strong sources, or for sources that show a significant spatial extent of spot emission. Hence, for all EVLA sources where the maser spots are concentrated in the center within the fitting uncertainties, only integrated spectra are shown in Fig.~\ref{evlaspectra}. We give a discussion of the spot morphologies for selected individual sources in Sect. \ref{indisources}.

In several sources with complex spectra, there were multiple maser spots for a given velocity channel. When these spots are close enough to be blended together, SAD does not provide a reliable fit. To overcome this problem, we inspected the fit results for each data cube. When multiple spots were blended in the fit results, we repeated the fits manually using the task ``JMFIT''. This procedure is adequate to reliably determine the maser spot morphologies for most sources. However, a small number of sources are too complex to resolve using this technique. These sources are noted in the discussion in Sect. \ref{indisources}. While Fig.~\ref{merlinspotmaps} shows the spot maps determined from our data, higher angular resolution observations are required to fully resolve individual spots in these sources and verify our results.

\subsection{Selected individual sources}\label{indisources}
{\noindent \it G35.03+0.35} -- A complex emission structure is seen in this source as expected from its line profile. There is an elongated though not perfectly linear structure seen between LSR velocities of 41.9 and 44.9~km~s$^{-1}$ with a linear velocity gradient across the structure. In addition, there are five other groups of spots covering the full velocity range of maser emission. The absolute position is consistent with that derived by \citet{cyga09} although their poorer resolution fails to resolve individual spots in channels with multiple spots. The 24~$\mu$m MIPSGAL image is saturated in the central pixels, and the overall morphology gives the appearance of two sources that are blended together. While Table~\ref{mipspos} gives the results of a single Gaussian fit, higher resolution data is required to determine the presence or absence of multiplicity. The source is associated with an extended green object (EGO) in the GLIMPSE data; a more detailed discussion can be found in \citet{cyga09}.

{\noindent \it G35.40+0.03} -- This is one of three sources for which no reliable counterpart can be determined at 24~$\mu$m. There is a point source 7\arcsec~away at $18^h55^m50^s.87$, $2\degr 12\arcmin 26\arcsec.4$ which is connected to extended emission towards the east suggestive of another point source at roughly $18^h55^m51^s.0$, $2\degr 12\arcmin 20\arcsec$ (no Gaussian fit could be obtained). The latter is about 3.4\arcsec~from the maser position. It is not clear whether either of the two sources is associated with the 6.7~GHz methanol maser.

{\noindent \it G35.79--0.17} -- This source was observed by \citet{bart09} and observed to have a linear morphology with a velocity gradient. While we detect this structure in our data, there are three other groups of spots, one of which has a ring shaped morphology. Moreover, the peak flux density detected in MERLIN is over 20~Jy, while \citet{bart09} detect a peak flux density of only 9.7~Jy. Hence, the EVN observations suffer from missing flux, and the overall morphology of the maser emission is complex rather than linear. We also note a significant position offset between the coordinates reported by \citet{bart09} and those in Table~\ref{merlinpos}. However, the position of \citet{bart09} is derived from MERLIN single baseline data which has a much larger uncertainty. The position offset in this source is consistent with the uncertainties quoted in \citet{szym07}.

{\noindent \it G36.70+0.09} -- The spot morphology as seen by MERLIN is very similar to that derived by \citet{bart09} using the EVN. There are two main groups of maser spots in this source. The southern group shows an inverted ``S'' structure, while the northern group hosts a linear distribution with a velocity gradient between 61.9 and 62.4~km~s$^{-1}$. The EVN observations, being more sensitive, also see more maser spots at high velocities ($\gtrsim 63$~km~s$^{-1}$), while we detect only a single spot at 63.0~km~s$^{-1}$.

{\noindent \it G36.84--0.02} -- There are two groups of maser spots in this source (Fig.~\ref{evlaspotmaps}) that are separated by 1.2\arcsec. However, both groups show spectral features within the overall velocity range containing emission. This suggests that the two groups are discrete sources rather than being kinematic features of a single source. However, given that the distance to the source is only 3.7 kpc, this would imply a physical separation of only 0.02 pc ($\sim 4500$ AU). High angular resolution mid-infrared and submillimeter observations are required to determine whether the two groups of spots are indeed discrete, or whether they are excited by a single source.

{\noindent \it G36.92+0.48} -- The MIPS 24~$\mu$m image shows a cluster of at least 2-3 sources within 10\arcsec~in this region. Taking into account the large distance to this source (15.8 kpc), the multiple 24~$\mu$m point sources probably reside in the same molecular cloud.

{\noindent \it G37.02--0.03 and G37.04--0.04} -- G37.04--0.04 is a new individual source detected in the MERLIN data, and is located 49\arcsec~away from G37.02--0.03. The spectral features of the source can be seen at a weak level in the published spectrum of G37.02+0.03 in Paper I, and was most probably missed in the original survey due to its close proximity to G37.02--0.03. The maser emission in G37.02+0.03 is confined to a small region, very similar to that observed with EVN by \citet{bart09}. We note however that the peak flux density as seen in MERLIN is much larger than that in EVN, and that maser spots outside the central core appear to be resolved out in the EVN data. In contrast, the morphology of G37.04--0.04 is more complex with more spectral features. There are two parallel emission features seen, each with a linear morphology. While the northern feature shows a linear velocity gradient, the southern feature does not show any coherent velocity structure and appears to be a mere superposition of different spectral features.

{\noindent \it G37.47--0.11} -- This is a complex source with several spectral features. The emission features between 60.8 and 63.1~km~s$^{-1}$ form a linear structure with a velocity gradient. However, the other spectral features are distributed randomly with a general tendency of features at lower velocity to be located to the south. The overall morphology is similar to that derived by \citet{cyga09} using the EVLA although we do not detect the eastern spots seen in their map. Consequently, we do not see the double arc structure surmised by \citet{cyga09}. It should also be noted that some channels showed emission in multiple spots which were unresolved by \citet{cyga09}.

{\noindent \it G37.53--0.11} -- There are two groups of maser spots in this source separated by 1.4\arcsec. At a distance of 9.9 kpc, this corresponds to a physical separation of 0.07 pc. High angular resolution observations at infrared or submillimeter wavelengths are required to determine whether the two groups are excited by a single or by two separate massive young stellar objects. The MIPSGAL image of this source is completely saturated and hence the position of the 24~$\mu$m point source cannot be measured.

{\noindent \it G37.55+0.19} -- There are two well separated groups of spectral features in this source which are also spatially separated in the spot map (Fig.~3). The features at lower velocity (78.4 to 80.0~km~s$^{-1}$) are situated about 0.2\arcsec~to the west of the other spots. The eastern group of spots is clustered in accordance with individual spectral features, but show a broad velocity trend from the north-west to the south-east with increasing velocity. As in G37.53--0.11, the position of the MIPSGAL counterpart cannot be determined due to saturation.

{\noindent \it G37.60+0.42} -- The intensity in our MERLIN spectrum is more than a factor of two higher than the EVN spectrum of \citet{bart09}. The peak emission in the MERLIN data occurs at 87.0~km~s$^{-1}$, while the EVN peak occurs at 85.8~km~s$^{-1}$. Since the masing spot at 87.0~km~s$^{-1}$ is about 30 mas to the east and 10 mas to the north of the spot at 85.8~km~s$^{-1}$, the reference coordinate quoted in Table~\ref{merlinpos} is different from that of \citet{bart09}. While the spot morphology determined from our data shows similarities to that derived from EVN, some differences exist. The MERLIN spots are much more tightly concentrated in declination between 84.9 and 89.0~km~s$^{-1}$, forming two linear structures with monotonic velocity gradients. The two easternmost maser spots of \citet{bart09} at velocities lower than 85~km~s$^{-1}$ are not detected in our data. However, we detect several spots at velocities greater than 91~km~s$^{-1}$ which are not detected in the EVN data. Another qualitative difference is that the spot distribution between 52 and 56~km~s$^{-1}$ resembles a ring rather than the linear structure seen in \citet{bart09}.

{\noindent \it G37.74--0.12} -- The MIPSGAL image of this source shows significant extended emission with the background subtracted image revealing a bow-shock morphology. It is not clear whether the morphology is intrinsic or whether it is due to a superposition of multiple sources. The 6.7~GHz methanol maser is coincident with the brightest emission region. The position reported in Table~\ref{mipspos} is the result of a Gaussian fit to the bright emission region.

{\noindent \it G37.76--0.19} -- The morphology of this source as seen with the EVLA shows a compact group of spots between 54.5 and 55.1~km~s$^{-1}$ (which is the strongest spectral feature in the source), while the other spectral features are distributed over a larger area to the east. The latter being weak, have relatively large fitting uncertainties, and so we do not attempt to deduce any morphologies from the data.

{\noindent \it G37.77--0.22} -- This methanol maser is embedded in a large area of extended 24~$\mu$m emission, and no point source can be discerned at the maser position.

{\noindent \it G38.03--0.30} -- As with G37.60, the MERLIN spectrum is much stronger than the EVN spectrum of \citet{bart09}, and the velocity of peak emission in the MERLIN data (58.2~km~s$^{-1}$) is different from that in the EVN data (55.7~km~s$^{-1}$). This leads to a small offset between the position quoted in Table~\ref{merlinpos} and that of \citet{bart09}. However, the spot morphology determined from MERLIN is consistent with (and essentially identical to) that determined using EVN, except for the detection of spots at high LSR velocities (62.8 and 63.7~km~s$^{-1}$) in the MERLIN data.

{\noindent \it G38.12--0.24} -- There are three groups of spots in this source. The velocity components between 68.1 and 71.4~km~s$^{-1}$ are tightly concentrated, while the two other groups of features, 76.7 -- 77.7~km~s$^{-1}$ and 79.1 -- 79.3~km~s$^{-1}$, lie to the southeast and northeast respectively. The weaker spectral feature around 74~km~s$^{-1}$, detected in only one channel, lies close to the central cluster of spots.

{\noindent \it G38.20--0.08} -- This is a complex source with several strong spectral features. We measure a peak flux density of 9.7~Jy at 84.3~km~s$^{-1}$, while the EVN data of \citet{bart09} detect a peak of only 0.83~Jy at 79.6~km~s$^{-1}$. Paper I lists a single dish peak flux density of 11.1~Jy at 79.6~km~s$^{-1}$. The MERLIN spectrum is very similar to that obtained at Arecibo except that the 84~km~s$^{-1}$ feature is much stronger in the MERLIN observation. While this can be attributed to variability, the significant discrepancy between EVN and MERLIN is unlikely to be due to variability, but suggests missing flux. The spot distribution is extremely complex with some channels showing as many as 3 spots some of which are blended together. Hence, the spot map shown in Fig.~\ref{merlinspotmaps} should be treated with caution. While most of the spots of \citet{bart09} can be identified in our spot map, we also detect several features that are not seen in EVN. We also do not detect any maser spot $\sim 180$ mas north-west of the central concentration of spots that is seen in the EVN data. The absolute position in Table~\ref{merlinpos} is consistent with that of \citet{bart09} when the different velocity of peak emission is taken into account.

The 24~$\mu$m MIPSGAL image shows a strong point source at $19^h01^m18^s.86$, $4\degr39\arcmin26\arcsec.7$ with the methanol maser being coincident with a weak source which lies on the Airy diffraction ring of the strong source. The weak source, which cannot be fit by a Gaussian, has an approximate position of $19^h01^m18^s.7$, $4\degr 39\arcmin 37\arcsec$ by visual inspection. Further work including PSF subtraction of the strong source is required to determine the position of the counterpart accurately. The position of the MIPS counterpart reported in Table~\ref{mipspos} is obtained by visual inspection.

{\noindent \it G38.26--0.08} -- This source (observed with EVLA) has three groups of spectral features which are spatially resolved. The features 6.3 -- 7.6~km~s$^{-1}$ and 11.7 -- 12.9~km~s$^{-1}$ populate the eastern and south-eastern part of the spot map respectively, while the strong spectral features between 14.2 and 15.5~km~s$^{-1}$ are more compact and lie to the west.

{\noindent \it G40.28--0.22} -- This is another complex source with an unusually large number of spectral features. The frequency used to observe this source led to the loss of features at velocities greater than $\sim 79$~km~s$^{-1}$. The overall shape of the spectrum obtained using MERLIN is very similar to that in Paper I although variability is seen in some spectral features -- the feature around 65~km~s$^{-1}$ is stronger while the peak flux density is lower in the MERLIN data compared to Paper I. The spot map shows about 8 groups of spots, at least two of which are extended linearly with a monotonic velocity gradient. However, the overall distribution of spots is complex with no velocity structure. The 24~$\mu$m MIPSGAL image is saturated in the center.

{\noindent \it G40.62--0.14} -- There are four spectral features in this source that are clearly resolved spatially. However, the overall morphology should be classified as ``multiple'' if not complex. The MIPSGAL 24~$\mu$m counterpart is fully saturated and hence its position could not be measured.

{\noindent \it G41.12--0.22} -- There are two spectral features in this source with the stronger feature showing a linear morphology with a velocity gradient.

{\noindent \it G41.23--0.20} -- There are four groups of spots in this source. While three groups show linear morphologies, two out of three groups do not show any clear velocity gradients. The MIPSGAL 24~$\mu$m image shows extended emission at the maser location. The background subtracted image shows a few point sources along with some residual extended emission. Hence there is some uncertainty in the position of the counterpart reported in Table~\ref{mipspos}.

{\noindent \it G41.34--0.14} -- The maser spots in this source fall into two broad groups. The spots at velocities less than 10~km~s$^{-1}$ lie to the southeast, while those at velocities greater than 10~km~s$^{-1}$ lie to the northwest. No distinct morphology can be discerned in the northwest, while the southeast group could possibly be fit with a ring morphology.

{\noindent \it G42.03+0.19} -- This is a complex source. The emission between 7.2 and 11.0~km~s$^{-1}$ are distributed in four clusters with an overall north-south trend with decreasing velocity. The other spots are distributed in a complex morphology though it is conceivable to fit the spots between 7.2 and 14.6~km~s$^{-1}$ with a ring distribution. We also detect a maser spot at 17.0~km~s$^{-1}$ although the MERLIN spectrum does not show a spectral feature at this velocity (this spectral feature is clearly seen in the Arecibo spectrum in Paper I). The position of the 24~$\mu$m counterpart has some uncertainty since a significant fraction of the PSF core is saturated and blanked.

{\noindent \it G42.43--0.26} -- The position of the 24~$\mu$m counterpart of this maser cannot be measured on account of it being completely saturated.

{\noindent \it G42.70--0.15} -- There are five spectral features in this source each of which are spatially distinct. While most individual features display a linear or arched morphology, there is no overall velocity structure in this source. It is possible that the different maser spots can be fit with a ring morphology.

{\noindent \it G43.04--0.46} -- There are three regions of emission in this source, two in the southeast and one in the northwest. The northwest group is offset from the other two groups by $\sim 2.5\arcsec$, which corresponds to a physical separation of 0.1 pc. While it is possible that the northwest group of spots is a distinct maser excited by a different YSO than that pumping the southeast group, complementary data at high angular resolution is required to confirm this scenario. The MIPSGAL counterpart is close to the southeast group of spots.

{\noindent \it W49N region} -- Using the MERLIN data, we are able to obtain accurate astrometry for all five methanol masers identified in Paper I. We do not identify any new emission regions to a 1$\sigma$ limit of 60 mJy beam$^{-1}$. Four out of five sources have previously published positions accurate to 0.4\arcsec~using the Australia Telescope Compact Array (ATCA; \citealt{casw09}). The two weak sources, G43.17--0.00 and G43.18--0.01 show a simple compact morphology or are detected in only one maser spot. The other three sources are discussed in more detail below. The entire region is saturated in the MIPSGAL 24~$\mu$m image and hence no 24~$\mu$m counterparts can be determined for any of these sources.

{\noindent \it G43.15+0.02} -- Although the spectrum in Fig.~\ref{merlinspotmaps} is affected by negative sidelobes of G43.16+0.02, the spot map was verified to not suffer from these effects. The spot map shows a simple linear morphology with an overall velocity gradient.

{\noindent \it G43.16+0.02} -- The spectral features between 7.7 and 8.9~km~s$^{-1}$, and 9.1 and 10.2 ~km~s$^{-1}$ form two roughly linear features with monotonic velocity gradients (though a sub-feature around 8.9~km~s$^{-1}$ is an outlier) that are perpendicular to each other. The emission between 15 and 20~km~s$^{-1}$ traces a linear and ring shaped morphology in addition to a few relatively isolated spots.

{\noindent \it G43.17+0.01} -- There are at least five morphological features in this source. The emission between 18.8 and 19.3~km~s$^{-1}$, and 20.0 and 20.5~km~s$^{-1}$ traces two linear structures in the northwest, while the 20.2 -- 20.7~km~s$^{-1}$ and 21.2 -- 22.1~km~s$^{-1}$ features form two clusters to the southeast. The emission between 20.2 and 21.1~km~s$^{-1}$ traces another linear feature which connects to the cluster between 21.2 and 22.1~km~s$^{-1}$. In addition, a number of spots between 19.1 and 20.0~km~s$^{-1}$ form a relatively random distribution.

{\noindent \it G43.80--0.13} -- This is a complex source with several maser spots being blended in our MERLIN data. While the spot distribution in Fig.~\ref{merlinspotmaps} shows two linear features between 39 and 41~km~s$^{-1}$, higher angular resolution data is required to confirm this. We also detect four other clusters of spots in this source. The spectrum in Fig.~\ref{merlinspotmaps} is Hanning smoothed to suppress ringing arising from the Gibbs phenomenon. The peak flux density is 53.9~Jy in the unsmoothed spectrum which reduces to 40.8~Jy after Hanning smoothing. Considering the velocity resolution of the MERLIN data, this source is consistent with no variability seen between the epochs of the Arecibo and MERLIN observations. The position of the maser's 24~$\mu$m counterpart cannot be measured on account of saturation.

{\noindent \it G44.31+0.04} -- The 24~$\mu$m counterpart of this maser is saturated in the central pixels.

{\noindent \it G45.07+0.13} -- Even though the peak flux density is over 60~Jy in this source, there are only two spectral features with the strong feature displaying a linear morphology with a monotonic velocity gradient. The MIPSGAL 24~$\mu$m image is fully saturated at the location of the maser, and so no point source counterpart could be identified.

{\noindent \it G45.44+0.07} -- This source is in a bright region with extended emission at 24~$\mu$m that is saturated close to the maser position. Although no point source is seen after background subtraction, this could be due to the close proximity to blanked region.

{\noindent \it G45.47+0.13} -- There are four emission regions in this source. Of particular interest is the double emission structure at 59.6 and 59.7~km~s$^{-1}$~(i.e. each velocity channel has emission in two spots) in the northeast. As with G45.44+0.07, this maser is located at the edge of a saturated region at 24~$\mu$m, and hence no 24~$\mu$m can be identified for the source.

{\noindent \it G45.47+0.05} -- While there are multiple spectral features in this source, the maser spots between 55.6 and 57.6~km~s$^{-1}$ form a single morphological feature although the velocity gradient is not monotonic across the spots between 56.5 and 57.6~km~s$^{-1}$. A second linear feature between 57.7 and 58.3~km~s$^{-1}$ is located about 10 mas to the east.

{\noindent \it G45.49+0.13} -- The maser spots in this source are concentrated in a compact region. The spot morphology is linear with a velocity gradient, except for the spot at 58.0~km~s$^{-1}$.

{\noindent \it G45.81--0.36} -- There are five emission regions in this source. The spectral features between velocities of 59.7 and 66.4~km~s$^{-1}$, which include the strongest peak, each display a simple morphology. In contrast, the features between 67.4 and 70.3~km~s$^{-1}$ show a more complex distribution with no overall velocity structure.

{\noindent \it G46.12+0.38} -- This source displays three spectral features, two of which have linear morphologies. The third feature is identified in only two spots, and hence no morphological information can be deduced for this feature.

{\noindent \it G48.90--0.27} -- While the overall morphology in this source is linear, there is no velocity structure across the feature.

{\noindent \it G48.99--0.30} -- This maser, as with a number of other sources in the W51 region is in a region that is saturated at 24~$\mu$m and hence no point source counterpart can be identified at this wavelength.

{\noindent \it G49.27+0.31} -- There are two emission regions in this source, one to the northeast and the other to the southwest. The southwest region shows a single extended morphology between velocities of --6.1 and --1.9~km~s$^{-1}$, and a monotonic velocity gradient is seen between --5.6 and --3.0~km~s$^{-1}$.

{\noindent \it G49.35+0.41} -- The emission in this maser is relatively compact with the spots between velocities of 66.9 and 69.1~km~s$^{-1}$ forming a reversed ``S'' shaped morphology.

{\noindent \it G49.41+0.33 \& G49.42+0.32} -- The single dish spectrum shows two main groups of spectral features, one between --27 and --23.5~km~s$^{-1}$, and a second group between -16 and --9.5~km~s$^{-1}$, although Paper I also shows weak emission around --19.5~km~s$^{-1}$. As indicated by \citet{cyga09}, at high angular resolution, the two groups of features separate into two distinct regions that are separated by $\sim 8\arcsec$ (0.47 pc physical separation at a distance of 12.2 kpc). The two sources are hence treated as distinct masers. Within our 1$\sigma$ limit of 45 mJy beam$^{-1}$, the western source (G49.41+0.33) only shows emission between --16 and --10~km~s$^{-1}$, while the eastern source (G49.42+0.32) only shows emission between --27 and --23.5~km~s$^{-1}$. We thus do not detect any maser spots at velocities less than --23~km~s$^{-1}$ in G49.41+0.33 as was seen by \citet{cyga09}. The high angular resolution of MERLIN clearly resolves the morphologies of both sources. G49.41+0.33 consists of a set of four morphological features extending in the east-west direction. Each feature is extended in the north-south direction as shown in Fig.~\ref{merlinspotmaps}. In contrast, G49.42+0.32 consists of five groups of maser spots spanning 140 mas in the north-south direction.

While G49.42+0.32 has a clear 24~$\mu$m counterpart, the case of G49.41+0.33 is ambiguous. There appears to be multiple point sources in the Airy ring of the source associated with G49.42+0.32, one of which is about 3\arcsec~away from G49.41+0.33 (coordinates of $19^h20^m59^s.0, 14\degr 46\arcmin 52\arcsec$ by visual inspection). Although it is not clear whether G49.41+0.33 has a 24~$\mu$m counterpart, it is associated with an EGO \citep{cyga09}. Higher spatial resolution mid-infrared observations are required to determine the counterpart of G49.41+0.33.

{\noindent \it W51 main} -- There are four methanol masers in the W51 main region identified in Paper I. This region was not observed in the present observing program. However, we have used the MERLIN data of \citet{xu09} to determine the positions of each of the four sources in the channels of strongest emission (see Table~\ref{merlinpos}). The dynamic range of the data is limited by the very strong source G49.49--0.39 (W51e2 in \citealt{xu09}). Moreover, the velocity resolution of this data is only 0.72~km~s$^{-1}$ after Hanning smoothing (required to reduce the Gibbs ringing from G49.49--0.39). Hence, no spot maps are derived for any of the sources. Our positions are in good agreement with those determined by \citet{casw09} using ATCA. \citet{casw09} also identified an additional site of emission in this region. Several regions of emission were also detected by \citet{phil05} using the EVN, although they do not provide any absolute positions for the sources. The entire region is saturated at 24~$\mu$m and hence no 24~$\mu$m counterparts can be determined for any of the sources.

{\noindent \it G49.60--0.25} -- This is an extremely complex source with several emission channels having multiple maser spots that are blended together. Although the spot map in Fig.~\ref{merlinspotmaps} is produced by manually fitting synthesized beams to the blended spots, observations at higher angular resolution are required to verify the same. The emission morphology is also extremely complex. The MIPSGAL 24~$\mu$m image suggests the presence of two sources that are blended together, with another source located on the Airy ring. The counterpart position indicated in Table~\ref{mipspos} was obtained by fitting two Gaussians to the central source(s).

{\noindent \it G52.92+0.41} -- This is an interesting source with three spectral features forming a linear north-south morphology. In addition, each spectral feature also shows a north-south morphology with a velocity gradient. However, there is no overall velocity gradient across the three features. The 24~$\mu$m counterpart is saturated at the central pixels.

{\noindent \it G53.04+0.11} -- The 24~$\mu$m image shows a cluster of at least four sources with the central two sources blended with each other. The counterpart listed in Table~\ref{mipspos} was obtained by fitting two Gaussians to the central pair of sources.

{\noindent \it G53.14+0.07} -- No 24~$\mu$m counterpart could be identified for this maser due to it being saturated.

{\noindent \it G53.62+0.04} -- The central pixels of the maser counterpart in the 24~$\mu$m MIPSGAL image are saturated.

\section{Discussion}

\subsection{Comparison with previous work}
Of the 82 sources with good astrometry, six (G35.79--0.17, G36.80+0.09, G37.02--0.03, G37.47--0.11, G38.03--0.30 and G38.20--0.08) have been studied using the EVN by \citet{bart09}, five (G35.03+0.35, G37.47--0.11, G37.53--0.11, G49.41+0.33 and G49.42+0.32) using the EVLA by \citet{cyga09}, and fourteen using the ATCA by \citet{casw09} (see Table 1 of \citealt{casw09}). In addition, the absolute positions of G43.15+0.02 and G49.41+0.33 (IRAS 19186+1440), derived from MERLIN (but at poorer velocity resolution) have been published by \citet{xu09}. For the EVN sources, most of our positions agree with previous work within our MERLIN astrometric uncertainty of 15 mas. The exceptions are G35.79--0.17 and G37.47--0.11 which have position differences of 140 mas and 270 mas respectively. However, the positions for these two sources were determined from MERLIN single baseline data (no EVN positions derived; see Sect. 3.1 of \citealt{bart09}) and have uncertainties of 175 mas and 365 mas respectively \citep{szym07}. The EVLA data of \citet{cyga09} were taken in the B configuration, for which uncertainties larger than 100 mas can be expected for the absolute astrometry. Taking this into consideration, there is good agreement between the positions in Table~\ref{merlinpos} and those of \citet{cyga09}. The positions derived from ATCA have an root mean square uncertainty of 0.4\arcsec~\citep{casw09}. Our positions in Table~\ref{merlinpos} agree well with those of \citet{casw09} except for G43.17--0.00 and 43.80--0.13. The discrepancy in the case of G43.17--0.00 is due to a typing mistake in the right ascension of \citet{casw09} (the right ascension should be $19^h10^m16^s.72$ instead of $19^h10^m16^s.27$; Caswell, private communication). There is also excellent agreement with the positions of \citet{xu09}, with minor differences presumably arising from the poorer velocity resolution of their data (which can lead to blended spots).

Comparing the positions in Tables~\ref{merlinpos} and \ref{evlapos} with those of Paper I, we find a mean offset between the accurate position and the single dish position (using a Gaussian probability density function) to be --8.5\arcsec~(--$0^s.56$) in right ascension and 1.2\arcsec~in declination. This confirms the observation of a bias in the right ascension of the Arecibo positions by \citet{casw09}. However, the cause of this offset is not clear. It is unlikely to be due to telescope pointing since no bias is seen in the pointing scans in the original data.

\subsection{Spatial extent of maser emission}

When analyzing the spatial extent of maser emission, we exclude sources that have suspected multiple YSOs exciting well separated spatial groups of spots (G36.84--0.02, G37.53--0.11 and G43.04--0.46) in addition to sources that have fewer than four maser spots. We derived spatial extents using the following procedure. We first defined the extent of the maser spot distribution in right ascension and declination, and carried out a coordinate rotation to determine an axis along with the extent was minimum. This was defined to be the minor axis, with the perpendicular axis being taken as the major axis. The spatial extent along the major axis ranges from 26 AU to 7680 AU with a median value of 775 AU, while the minor axis ranges from 2 AU to 2400 AU with a median value of 180 AU. Fig.~\ref{majexthist} shows the distribution of the spatial extent along the major axis. It can be seen that the majority of the sources have spot distributions with major axis less than 1000 AU.

It is of interest to look for correlations between the spatial major axis extent of maser emission (the spatial extent along the minor axis is affected by the three-dimensional geometry of the source) and other parameters such as the velocity width of emission and maser luminosity. Fig. \ref{sizevelcorr} shows the spatial extent as a function of the range in velocity of maser emission (i.e. $\Delta V = V_{max} - V_{min}$ where $V_{min}$ and $V_{max}$ are the minimum and maximum velocities of maser emission respectively) along with a linear fit to the data. While there is significant scatter, it can be seen that sources that have a larger velocity width tend to have spots distributed over a larger spatial extent. Fig. \ref{sizevelcorr} clearly shows that the minimum value of the spatial major axis extent of maser spots (for a given maser velocity width) has a strong dependence on the velocity width of maser emission.

Fig. \ref{sizelumcorr} shows the spatial major axis extent of maser emission as a function of maser luminosity (taken from Paper III). A power law fit to the data (dashed line in Fig. \ref{sizelumcorr}) gives an exponent of 0.51. However, as in Fig. \ref{sizevelcorr}, there is significant scatter in this relationship. It would be of particular interest to look for correlations between the spatial extent of maser emission and the luminosity of the central YSO that excites the maser. However, this requires determination of spectral energy distributions of the central object from millimeter to mid-infrared wavelengths, which we reserve for a future paper.

\subsection{Methanol maser morphologies}

Figs. \ref{merlinspotmaps} and \ref{evlaspotmaps} make it clear that it is difficult to classify methanol maser morphologies. However, there are a few broad classes: (i) {\it Linear/arched} where the overall morphology forms a rough straight line or a smooth curve although there may not be any systematic velocity gradient; (ii) {\it Compact/simple} morphologies where the extent of the maser spots is less than $\sim 20$ mas; (iii) {\it Multiple}, where different spectral features are well separated into compact clusters of maser spots (the {\it pair} morphology used in previous studies would be a special case of this class); and (iv) {\it Complex} where no regular structure is seen. 

Using the classification scheme above, there are 9 sources with linear/arched morphology, 5 with compact/simple morphology, 12 with multiple groups and 24 with complex morphology (see Tables \ref{merlinpos} and \ref{evlapos}). It is remarkable that we do not see any source with a clear ring morphology although there are a few sources (G35.79--0.17, G36.70+0.09, G41.34--0.14, G42.03+0.19, G42.70--0.15 and G43.16+0.02) where there are potential ring structures. However, these structures are usually embedded in a larger more complex distribution of maser spots. In contrast, \citet{bart09} found 29\% of their sample to have ring like emission morphologies. The latter may be a selection effect due to their very high spatial resolution resolving a significant fraction of the overall emission. For the five sources in common for which emission morphologies have been determined, four have missing flux between 70 and 90\%. Furthermore, a comparison between the single dish fluxes of \citet{szym00,szym02} and the EVN fluxes of \citet{bart09} shows a similar trend for a large number of sources. 

The large fraction of missing flux in some sources indicates the presence of emission structures on scales that are resolved by the EVN. This is most likely to arise from the structure of the maser spot itself. For example, \citet{mini02} found a majority of masing spots to consist of a compact core with a halo of extended emission that is resolved even on short VLBI baselines. However, if some spots do not have a compact core, they would not be detected in VLBI observations. The comparison between the overall emission morphology in Fig.~\ref{merlinspotmaps} and that of \citet{bart09} suggests examples for both scenarios. For example, the spot morphology in G37.02--0.03 obtained using MERLIN is remarkably similar to that of EVN in spite of the latter missing 90\% of the flux, suggesting that most of the maser spots in this source have a core/halo structure. In contrast, G35.79--0.17 and G38.20--0.08 show several maser spots in the MERLIN data that are not seen using EVN suggesting that a number of spots in these sources do not have a compact core. Although a detailed discussion about the structure of a maser spot is beyond the scope of this paper and would require combining this data with higher resolution observations, the work of \citet{harv06} shows that in very nearby regions, even MERLIN observations could resolve some extended emission. We discuss this in the context of variability in Sect. \ref{variability}.

In light of this discussion, it appears likely that at least some of the ring shaped emission structures seen by \citet{bart09} are embedded in or are part of larger structures. This casts doubt on simple kinematic models for such sources, and highlights the dangers of interpreting kinematics solely based on morphology. To obtain a better understanding of the origin of methanol maser emission in the context of a massive young stellar object, one requires either proper motion studies through multi-epoch VLBI, or high angular resolution data in mid-infrared and submillimeter wavelengths.

\subsection{Variability of AMGPS methanol masers}\label{variability}
Methanol masers are known to be variable, and a number of studies have systematically studied the variability of the masers (e.g. \citealt{goed04}). To look at the variability of the AMGPS masers, we compared the integrated spectra from MERLIN and EVLA with those in Paper I. Among the 77 sources with spectra in both MERLIN/EVLA and Paper I, 23 have similar peak flux densities, 16 have become stronger and 38 are weaker. Among the variable sources, 20 sources have changes in their peak flux densities by more than 50\%, the most dramatic being G35.59+0.06 which has dropped from 0.82~Jy in the Arecibo data to 0.20~Jy in the EVLA data. It is interesting to note that all the masers in the W49N region have become weaker with three sources being weaker by more than 50\%.

While some MERLIN sources could suffer from resolving out diffuse emission (e.g. \citealt{harv06}), this is unlikely for the more distant sources. In any case, the poorer resolution of the EVLA is expected to recover all the flux in the source. Among the 25 EVLA sources, 5 have the same peak flux densities as Paper I, while 15 are weaker (although the change is small in three sources) and 5 are stronger. We however caution that this statistic has a bias towards sources that have become weaker since they are more likely to be not detected in the MERLIN data. Nonetheless, it appears that a larger fraction of the AMGPS masers have become weaker compared to the flux measurement in Paper I. It is interesting to note a similar trend in Paper I when compared with previous work (see Table~2 of Paper I).

\subsection{Association with mid-infrared emission}\label{mirassociation}
The association between mid-infrared emission and 6.7~GHz methanol masers has been explored in a number of previous works. While early studies found a number of sources to not have mid-infrared counterparts at 10.5~$\mu$m and/or 20.0~$\mu$m \citep{wals01} or in the MSX point source catalog (e.g. Paper II), recent work has found methanol masers to be generally associated with mid-infrared emission. For example, \citet{elli06} found most methanol masers to have a counterpart in the Spitzer GLIMPSE survey within 5\arcsec. The more recent work of \citet{xu09}, \citet{cyga09}, and \citet{bart09} have found very close correspondence between 6.7~GHz methanol masers and mid-infrared emission at 24~$\mu$m and 4.5 $\mu$m.

Fig.~\ref{mipssephist} shows the histogram of the angular physical separation between 6.7~GHz methanol masers and their 24~$\mu$m counterparts for the 60 sources in Table~\ref{mipspos}. The median angular separation between a methanol maser and its 24~$\mu$m counterpart is 0.9\arcsec~and the corresponding physical separation is 7100 AU. We also note that 18 sources had no counterpart identified due to detector saturation, and only four sources had no counterpart identified in the MIPSGAL data. Of the four, two were in regions of extended emission, and an additional source has a potential counterpart on the Airy ring of a stronger source. Hence, almost all 6.7~GHz methanol masers have associated mid-infrared emission. These results also show that the lack of counterparts in the MSX Galactic plane survey is primarily a result of poorer sensitivity and to a lesser extent the poorer resolution of the MSX satellite. The 24~$\mu$m point source fluxes of the methanol maser counterparts are frequently less than 1~Jy (see Table~3 of \citealt{pand10}) which would be undetected by MSX at 21 $\mu$m, while the poorer resolution of MSX would lead to significant confusion in some crowded fields.

This close association of mid-infrared point sources associated with 6.7 GHz methanol masers is important in the context of theoretical models for the maser emission. For a wavelength of 24~$\mu$m, the Planck function has its maximum at 120 K, which is very close to the temperature of the molecular hot cores that the high mass protostellar objects exciting class II methanol masers reside in. This wavelength is also close to that of one the main bands invoked in radiative class II methanol maser pumping schemes (excitation from the torsional ground state, $\upsilon_t = 0$ to $\upsilon_t = 2$; see \citealt{sobo94}). These two facts together, combined with the ubiquitous presence of 24~$\mu$m sources apparently ensure extremely effective excitation of the masers.

The work of \citet{bart09} found that masers with ring shaped and regular morphologies had strong 8~$\mu$m emission. Particularly, they found that no source that was saturated at 8~$\mu$m had a complex morphology. While we have not inspected the 8~$\mu$m images of the AMGPS sources, we do not see a similar clear correlation with the 24~$\mu$m data. Considering the 16 sources that were saturated at 24~$\mu$m (even if only in the central pixels) and had morphologies determined, there were 5 linear/arched, 3 multiple, one compact and 7 complex sources. There is no statistically significant difference between the morphologies of masers associated with sources that are saturated at 24~$\mu$m and that of the overall sample. However, further work is required to verify whether a similar trend is seen with the GLIMPSE 8~$\mu$m data.

It is interesting to note that from Fig.~\ref{mipssephist}, the separation between the maser and mid-infrared source does not peak at zero. This could arise either from the uncertainty in the absolute astrometry of MIPSGAL, or from astrophysical reasons underlying the location of the maser emission relative to the source of mid-infrared emission. \citet{care09} analyze the relative positions of 24~$\mu$m MIPSGAL point sources with those of 8~$\mu$m counterparts in the GLIMPSE point source catalog, and find the median astrometric offset between the MIPSGAL and GLIMPSE to be 0.85\arcsec. This is very close to the median separation between methanol masers and MIPSGAL counterparts, and may be the likely explanation for this phenomenon. We have not undertaken the exercise of cross-correlating the maser positions with the GLIMPSE catalog due to the presence of emission and absorption features from dust grains in some IRAC bands, which often leads to extended emission. Moreover, it is clear from the work of \citet{cyga09} that many infrared counterparts of methanol masers are also extended at 4.5 $\mu$m. We thus reserve addressing the relation between GLIMPSE point sources and the AMGPS masers for a future paper. However, the issue of whether or not there is a systematic offset between mid-infrared sources and methanol masers for astrophysical reasons can be resolved by high spatial resolution imaging from ground based facilities.

An interesting consequence of the close correspondence between MIPSGAL 24~$\mu$m point sources and methanol masers is that one can use the former to estimate positions for masers that have not been detected with an interferometer. In the AMGPS catalog, there are six sources which are not detected, or haven't been observed due to constraints of telescope time. Since the 1$\sigma$ uncertainty in the AMGPS position is 7\arcsec, we looked for 24~$\mu$m point sources within 18\arcsec~of each of the six sources. For five sources, there is a clear 24~$\mu$m counterpart, while for G36.90--0.41, there are two MIPSGAL point sources within 18\arcsec~of the Arecibo position. The coordinates of the 24~$\mu$m sources are listed in Table~\ref{otherpos}. Considering the statistics of the offset between MIPSGAL point sources and methanol masers, we suggest that these positions can be used to refine the single dish positions to an accuracy of $\sim 1\arcsec$. 

\section{Conclusions}

We have determined the absolute positions of 82 methanol masers in the region surveyed by the AMGPS including two sources that were not present in the original catalog. The median spatial extent along the major axis of maser emission is $\sim 775$~AU. The spatial extent is correlated to both the velocity width of the maser and the maser luminosity though there is significant scatter in both relationships. Among the 50 sources for which the morphologies of maser emission could be determined, 9 sources have a linear/arched morphology, 5 have compact/simple structures, 12 have multiple compact groups of emission and 24 have complex morphology. In sources where there is a large discrepancy between the flux recovered in VLBI experiments and that measured with a single dish, we often find the missing flux to be distributed over complex structures. Hence, some relatively simple morphological structures seen in VLBI are likely to be part of a more complex emission structure in at least some sources. This makes interpretation of the origin of maser emission difficult, and multi-epoch proper motion studies or complementary high resolution mid-infrared or submillimeter observations are required to resolve this problem. We also find very close correspondence between mid-infrared emission and 6.7~GHz methanol masers as would be expected from theoretical models of the maser emission.

\acknowledgements
We are extremely grateful to Anita Richards for invaluable help in MERLIN data processing and other useful discussions. We also thank Rob Beswick and Peter Thomasson for their assistance in MERLIN data reduction, Anna Bartkiewicz for useful discussion, and Jim Caswell whose comments helped improve the paper. YX was supported by the Chinese NSF through grants NSF 11073054, NSF10673024, NSF 10733030, NSF 10703010 and NSF 10621303, and NBRPC (973 Program) under grant 2007CB815403. This work has benefited from research funding from the European Community's sixth Framework Programme under RadioNet R113CT 2003 5058187. This work was carried out in part by the Jet Propulsion Laboratory, California Institute of Technology. This research has made use of NASA's Astrophysics Data System.

\begin{deluxetable}{cccccc}
\tabletypesize{\footnotesize}
\tablecaption{Positions of the phase calibrators used in the MERLIN observations. \label{merlincalib}}
\tablewidth{0pt}
\tablehead{
\colhead{Source name} & \colhead{RA (J2000)} & \colhead{Dec. (J2000)} & \colhead{$\sigma_\alpha$} & \colhead{$\sigma_\delta$} & \colhead{Reference} \\
\colhead{} & \colhead{(h m s)} & \colhead{(\degr~\arcmin~\arcsec)} & \colhead{(mas)} & \colhead{(mas)} & \colhead{}
}
\startdata
B1904+013 & 19 07 11.996253 & 01 27 08.96250 & 1.12 & 1.35 & 1 \\
B1854+061 & 18 56 31.838813 & 06 10 16.76649 & 0.62 & 0.87 & 2 \\
B1919+086 & 19 22 18.633756 & 08 41 57.37052 & 1.80 & 4.03 & 3 \\
B1920+154 & 19 22 34.699314 & 15 30 10.03262 & 0.73 & 0.78 & 4 \\
B1922+155 & 19 24 39.455870 & 15 40 43.94172 & 0.32 & 0.26 & 4 \\
\enddata
\tablecomments{The columns show the source name, equatorial J2000 coordinates, uncertainties in right ascension and declination, and the reference for the same.}
\tablerefs{
(1) Petrov et al. 2005; (2) Fomalont et al. 2003; (3) VLBA calibrator manual: http://www.vlba.nrao.edu/astro/calib/vlbaCalib.txt; (4) Beasley et al. 2002 
}
\end{deluxetable}

\begin{deluxetable}{cccccc}
\tabletypesize{\footnotesize}
\tablecaption{Positions of the phase calibrators used in the EVLA observations. \label{evlacalib}}
\tablewidth{0pt}
\tablehead{
\colhead{Source name} & \colhead{RA (J2000)} & \colhead{Dec. (J2000)} & \colhead{$\sigma_\alpha$} & \colhead{$\sigma_\delta$} & \colhead{Reference} \\
\colhead{} & \colhead{(h m s)} & \colhead{(\degr~\arcmin~\arcsec)} & \colhead{(mas)} & \colhead{(mas)} & \colhead{}
}
\startdata
J1851+005 & 18 51 46.723009 & 00 35 32.36604 & 1.82 & 2.30 & 1 \\
J1856+061 & 18 56 31.838813 & 06 10 16.76649 & 0.62 & 0.87 & 1 \\
J1922+155 & 19 22 34.699314 & 15 30 10.03262 & 0.73 & 0.78 & 2 \\
\enddata
\tablecomments{The columns show the source name, equatorial J2000 coordinates, uncertainties in right ascension and declination, and the reference for the same.}
\tablerefs{
(1) Fomalont et al. 2003; (2) Beasley et al. 2002
}
\end{deluxetable}

\begin{deluxetable}{ccccccccc}
\tabletypesize{\footnotesize}
\tablecaption{Absolute positions of the AMGPS 6.7~GHz methanol masers determined using MERLIN.\label{merlinpos}}
\tablewidth{0pt}
\tablehead{
\colhead{Source name} & \colhead{RA (J2000)} & \colhead{Dec. (J2000)} & \colhead {$l$} & \colhead{$b$} & \colhead{$S_p$} & \colhead{$V_p$} & \colhead{$V_{\mathrm{sys}}$} & \colhead{Morphology} \\
\colhead{} & \colhead{(h m s)} & \colhead{(\degr~\arcmin~\arcsec)} & \colhead{(\degr)} & \colhead{(\degr)} & \colhead{(Jy)} & \colhead{(km s$^{-1}$)} & \colhead{(km s$^{-1}$)} & \colhead{}
}
\startdata
G35.03+0.35  & 18 54 00.658 & 02 01 19.23 & 35.02469 &   0.34981 & 17.8 & 44.4 & 52.8 & C \\
G35.25--0.24 & 18 56 30.388 & 01 57 08.88 & 35.24726 & --0.23675 & 0.90 & 72.4 & 62.0 & L/A \\
G35.79--0.17 & 18 57 16.892 & 02 27 58.05 & 35.79267 & --0.17456 & 22.2 & 60.7 & 60.0 & M \\
G36.64--0.21 & 18 58 55.236 & 03 12 04.72 & 36.63362 & --0.20287 & 0.74 & 77.5 & 75.2 & \\
G36.70+0.09  & 18 57 59.123 & 03 24 06.12 & 36.70526 &   0.09630 &  7.5 & 53.1 & 59.7 & C \\
G36.92+0.48  & 18 56 59.786 & 03 46 03.60 & 36.91820 &   0.48293 & 0.47 & --35.9 & --30.7 \\
G37.02--0.03 & 18 59 03.642 & 03 37 45.08 & 37.03020 & --0.03847 &  6.9 & 78.4 & 80.6 & S \\
G37.04--0.04 & 18 59 04.406 & 03 38 32.77 & 37.04343 & --0.03524 &  6.0 & 84.7 & 80.6 & C \\
G37.47--0.11 & 19 00 07.144 & 03 59 53.08 & 37.47897 & --0.10467 &  5.8 & 58.3 & 58.8 & C \\
G37.53--0.11 & 19 00 16.056 & 04 03 16.09 & 37.54606 & --0.11182 &  2.5 & 50.1 & 53.0 & M \\
G37.60+0.42  & 18 58 26.799 & 04 20 45.47 & 37.59784 &   0.42522 & 16.3 & 87.0 & 89.5 & C \\
G38.03--0.30 & 19 01 50.470 & 04 24 18.94 & 38.03748 & --0.30019 &  8.3 & 58.2 & 62.0 & C \\
G38.12--0.24 & 19 01 44.152 & 04 30 37.42 & 38.11894 & --0.22873 &  3.0 & 70.7 & 83.2 & M \\
G38.20--0.08 & 19 01 18.730 & 04 39 34.32 & 38.20322 & --0.06660 &  9.7 & 84.3 & 83.3 & C \\
G38.66+0.08  & 19 01 35.244 & 05 07 47.36 & 38.65275 &   0.08767 &  1.3 & --31.3 & --39.2 & \\
G38.92--0.36 & 19 03 38.659 & 05 09 42.49 & 38.91576 & --0.35296 &  2.4 & 32.1 & 38.7 & S \\
G39.39--0.14 & 19 03 45.312 & 05 40 42.68 & 39.38764 & --0.14060 & 0.26 & 60.4 & 66.1 &  \\
G40.28--0.22 & 19 05 41.215 & 06 26 12.69 & 40.28179 & --0.21929 & 20.5 & 74.1 & 73.0 & C \\
G40.62--0.14 & 19 06 01.630 & 06 46 36.18 & 40.62249 & --0.13821 & 17.4 & 31.1 & 32.3 & M \\
G40.94--0.04 & 19 06 15.378 & 07 05 54.49 & 40.93438 & --0.04086 &  3.4 & 36.6 & 40.2 & M \\
G41.12--0.22 & 19 07 14.856 & 07 11 00.69 & 41.12299 & --0.22011 &  1.3 & 63.6 & 60.1 & M \\
G41.23--0.20 & 19 07 21.378 & 07 17 08.17 & 41.22603 & --0.19708 &  5.1 & 57.2 & 59.2 & M \\
G41.34--0.14 & 19 07 21.842 & 07 25 17.27 & 41.34753 & --0.13626 & 11.8 & 11.7 & 12.6 & C \\
G42.03+0.19  & 19 07 28.185 & 08 10 53.47 & 42.03438 &   0.19025 & 16.1 & 12.8 & 18.0 & C \\
G42.30--0.30 & 19 09 43.592 & 08 11 41.41 & 42.30354 & --0.29923 &  6.2 & 28.1 & 27.1 & C \\
G42.43--0.26 & 19 09 49.858 & 08 19 45.40 & 42.43474 & --0.26011 &  1.6 & 66.8 & 64.5 &  \\
G42.70--0.15 & 19 09 55.069 & 08 36 53.45 & 42.69798 & --0.14735 &  3.2 & --42.9 & --44.3 & C\\
G43.04--0.46 & 19 11 38.984 & 08 46 30.71 & 43.03797 & --0.45282 &  8.7 & 54.8 & 57.3 & M \\
G43.08--0.08 & 19 10 22.050 & 08 58 51.49 & 43.07404 & --0.07686 &  6.5 & 10.2 & 12.3 & C \\
G43.15+0.02  & 19 10 11.049 & 09 05 20.46 & 43.14897 &   0.01319 & 10.2\tablenotemark{a} & 13.3 & 10.9 & L/A \\
G43.16+0.02  & 19 10 12.883 & 09 06 12.22 & 43.16521 &   0.01314 & 20.1 &  9.3 &  3.5 & C \\
G43.17+0.01  & 19 10 15.356 & 09 06 15.50 & 43.17072 &   0.00453 &  5.9 & 20.2 & 12.0 & C \\
G43.17--0.00 & 19 10 16.720 & 09 05 51.27 & 43.16734 & --0.00355 &  1.6 & --1.1 &  1.7 & S \\
G43.18--0.01 & 19 10 20.068 & 09 05 55.87 & 43.17483 & --0.01518 & 0.32 & 11.2 & 11.9 & \\
G43.80--0.13 & 19 11 53.990 & 09 35 50.61 & 43.79542 & --0.12709 & 53.9\tablenotemark{b} & 39.6 & 43.7 & C \\
G45.07+0.13  & 19 13 22.129 & 10 50 53.11 & 45.07122 &   0.13213 & 63.5 & 57.8 & 59.2 & L/A \\
G45.44+0.07  & 19 14 18.291 & 11 08 58.97 & 45.44501 &   0.06868 & 0.68 & 50.0 & 58.5 &  \\
G45.47+0.13  & 19 14 07.362 & 11 12 15.98 & 45.47272 &   0.13366 &  8.4 & 65.7 & 61.9 & M \\
G45.47+0.05  & 19 14 24.147 & 11 09 43.43 & 45.46706 &   0.05321 &  5.7 & 56.0 & 59.7 & M \\
G45.49+0.13  & 19 14 11.357 & 11 13 06.41 & 45.49271 &   0.12570 &  9.7 & 57.3 & 60.5 & L/A \\
G45.81--0.36 & 19 16 31.081 & 11 16 12.01 & 45.80404 & --0.35574 & 14.3 & 59.9 & 60.3 & C \\
G46.12+0.38  & 19 14 25.520 & 11 53 25.99 & 46.11477 &   0.38670 & 0.90 & 59.0 & 55.0 & C \\
G48.90--0.27 & 19 22 10.330 & 14 02 43.51 & 48.90213 & --0.27331 & 0.52 & 72.0 & 68.4 & L/A \\
G48.99--0.30 & 19 22 26.134 & 14 06 39.78 & 48.99012 & --0.29871 & 0.74 & 71.5 & 67.5 \\
G49.27+0.31  & 19 20 44.859 & 14 38 26.91 & 49.26513 &   0.31130 &  8.2 & --4.5 &  3.4 & M \\
G49.35+0.41  & 19 20 32.449 & 14 45 45.44 & 49.34917 &   0.41268 & 11.0 & 68.0 & 65.5 & S \\
G49.41+0.33  & 19 20 59.211 & 14 46 49.66 & 49.41558 &   0.32588 & 10.1 & -12.1 & --21.3 & C \\
G49.42+0.32  & 19 20 59.817 & 14 46 49.09 & 49.41659 &   0.32365 &  3.1 & -26.0 & --21.3 & L/A \\
G49.47--0.37 & 19 23 37.897 & 14 29 59.26 & 49.46961 & --0.37061 &  2.9 & 64.1 & 62.4 & \\
G49.48--0.40 & 19 23 46.191 & 14 29 47.06 & 49.48244 & --0.40170 &  2.8 & 50.0 & 58.2 & \\
G49.49--0.37 & 19 23 39.824 & 14 31 04.94 & 49.48936 & --0.36884 & 18.9 & 56.4 & 60.6 & \\
G49.49--0.39 & 19 23 43.949 & 14 30 34.45 & 49.48977 & --0.38750 &  247 & 59.2 & 56.6 & \\
G49.60--0.25 & 19 23 26.611 & 14 40 16.99 & 49.59933 & --0.24941 & 48.6 & 63.1 & 57.2 & C \\
G50.78+0.15  & 19 24 17.411 & 15 54 01.60 & 50.77881 &   0.15196 &  4.7 & 49.1 & 42.2 & S \\
G52.92+0.41  & 19 27 34.960 & 17 54 38.14 & 52.92177 &   0.41404 &  4.9 & 39.1 & 44.8 & L/A \\
G53.14+0.07  & 19 29 17.581 & 17 56 23.21 & 53.14183 &   0.07062 & 0.73 & 24.6 & 21.8 & L/A \\
G53.62+0.04  & 19 30 23.016 & 18 20 26.68 & 53.61793 &   0.03553 & 11.2 & 19.0 & 23.3 & L/A \\
\enddata
\tablecomments{The columns show the source name as defined in Paper I (with a prefix `G' added to indicate derivation from Galactic coordinates), equatorial and Galactic coordinates, peak flux density, velocity of peak emission, the systemic velocity and the overall emission morphology. The positions are derived from the channel with peak emission and serve as the reference coordinate for the spot maps of Fig.~\ref{merlinspotmaps}. The morphology classes are linear/arched (L/A), compact/simple (S), multiple (M) and complex (C).}
\tablenotetext{a}{Peak flux density is 9.4~Jy after Hanning smoothing.}
\tablenotetext{b}{Peak flux density is 40.8~Jy after Hanning smoothing.}
\end{deluxetable}

\begin{deluxetable}{ccccccccc}
\tabletypesize{\footnotesize}
\tablecaption{Absolute positions of the AMGPS 6.7~GHz methanol masers determined using EVLA.\label{evlapos}}
\tablewidth{0pt}
\tablehead{
\colhead{Source name} & \colhead{RA (J2000)} & \colhead{Dec. (J2000)} & \colhead {$l$} & \colhead{$b$} & \colhead{$S_p$} & \colhead{$V_p$} & \colhead{$V_{\mathrm{sys}}$} &\colhead{Morphology}  \\
\colhead{} & \colhead{(h m s)} & \colhead{(\degr~\arcmin~\arcsec)} & \colhead{(\degr)} & \colhead{(\degr)} & \colhead{(Jy)} & \colhead{(km s$^{-1}$)} & \colhead{(km s$^{-1}$)} & \colhead{}
}
\startdata
G35.40+0.03  & 18 55 50.779 & 02 12 19.08 & 35.39697 &   0.02538 & 0.66 & 89.2 & 95.0 & \\
G35.59+0.06  & 18 56 04.219 & 02 23 28.34 & 35.58792 &   0.06043 & 0.20 & 44.1 & 49.1 & \\
G36.02--0.20 & 18 57 45.868 & 02 39 05.67 & 36.01273 & --0.19719 & 0.06 & 92.6 & 87.3 & \\
G36.84--0.02 & 18 58 39.214 & 03 28 00.89 & 36.83944 & --0.02224 &  2.1 & 61.7 & 59.2 & M \\
G37.38--0.09 & 18 59 51.586 & 03 55 18.02 & 37.38145 & --0.08208 & 0.06 & 70.7 & 57.7 & \\
G37.55+0.19  & 18 59 09.985 & 04 12 15.54 & 37.55382 &   0.20089 &  4.5 & 83.7 & 85.3 & C \\
G37.74--0.12 & 19 00 36.841 & 04 13 19.98 & 37.73474 & --0.11193 & 0.89 & 50.3 & 45.5 & \\
G37.76--0.19 & 19 00 55.421 & 04 12 12.56 & 37.75340 & --0.18915 &  2.1 & 54.6 & 59.2 & C \\
G37.77--0.22 & 19 01 02.268 & 04 12 16.55 & 37.76740 & --0.21394 & 0.49 & 69.6 & 61.9 & \\
G38.08--0.27 & 19 01 47.317 & 04 27 20.90 & 38.07642 & --0.26541 & 0.54 & 67.6 & 64.6 & \\
G38.26--0.08 & 19 01 26.233 & 04 42 17.26 & 38.25772 & --0.07359 &  8.1 & 15.4 & 11.6 & C \\
G38.26--0.20 & 19 01 52.956 & 04 38 39.47 & 38.25473 & --0.19995 & 0.58 & 70.2 & 65.4 & \\
G38.56+0.15  & 19 01 08.345 & 05 04 36.71 & 38.55457 &   0.16269 & 0.44 & 31.5 & 29.3 & \\
G38.60--0.21 & 19 02 33.461 & 04 56 36.37 & 38.59767 & --0.21249 & 0.26 & 62.5 & 66.4 & \\
G41.08--0.13 & 19 06 49.047 & 07 11 06.57 & 41.07536 & --0.12462 & 0.17 & 57.6 & 63.8 & \\
G41.12--0.11 & 19 06 50.248 & 07 14 01.49 & 41.12079 & --0.10668 & 0.43 & 36.5 & 38.1 & \\
G41.16--0.20 & 19 07 14.369 & 07 13 18.08 & 41.15595 & --0.20076 &  1.3 & 56.0 & 59.9 & C \\
G41.27+0.37  & 19 05 23.606 & 07 35 05.25 & 41.26807 &   0.37259 & 0.22 & 20.4 & 14.6 & \\
G41.58+0.04  & 19 07 09.178 & 07 42 25.24 & 41.57699 &   0.04156 & 0.38 & 12.0 & 12.4 & \\
G44.31+0.04  & 19 12 15.816 & 10 07 53.52 & 44.31035 &   0.04078 & 0.54 & 55.7 & 57.0 & \\
G44.64--0.52 & 19 14 53.766 & 10 10 07.69 & 44.64395 & --0.51580 & 0.81 & 49.5 & 46.0 & \\
G45.57--0.12 & 19 15 13.152 & 11 10 16.54 & 45.56834 & --0.11987 & 0.22 &  1.5 &  4.8 & \\
G46.07+0.22  & 19 14 56.077 & 11 46 12.98 & 46.06619 &   0.22046 & 0.81 & 23.6 & 19.5 & \\
G49.62--0.36 & 19 23 52.805 & 14 38 03.25 & 49.61650 & --0.36002 & 0.16 & 50.2 & 54.9 & \\
G53.04+0.11  & 19 28 55.494 & 17 52 03.11 & 53.03648 &   0.11292 &  1.6 & 10.1 &  5.5 & \\
\enddata
\tablecomments{The columns show the source name as defined in Paper I (with a prefix `G' added to indicate derivation from Galactic coordinates), equatorial and Galactic coordinates, peak flux density, velocity of peak emission and the systemic velocity. The positions are derived from the channel with peak emission and serve as the reference coordinate for the spot maps of Fig.~\ref{evlaspotmaps}. The morphology classes are linear/arched (L/A), compact/simple (S), multiple (M) and complex (C).}
\end{deluxetable}

\begin{deluxetable}{ccccccc}
\tabletypesize{\footnotesize}
\tablecaption{MIPS 24~$\mu$m counterparts of the AMGPS 6.7~GHz methanol masers.\label{mipspos}}
\tablewidth{0pt}
\tablehead{
\colhead{Source name} & \colhead{RA (J2000)} & \colhead{Dec. (J2000)} & \colhead {} & \colhead{Source name} & \colhead{RA (J2000)} & \colhead{Dec. (J2000)} \\
\colhead{} & \colhead{(h m s)} & \colhead{(\degr~\arcmin~\arcsec)} & \colhead{} & \colhead{} & \colhead{(h m s)} & \colhead{(\degr~\arcmin~\arcsec)}
}
\startdata
G35.03+0.35  & 18 54 00.64 & 02 01 21.0 & & G41.12--0.11 & 19 06 50.20 & 07 14 02.3 \\
G35.25--0.24 & 18 56 30.36 & 01 57 09.6 & & G41.12--0.22 & 19 07 14.77 & 07 11 01.6 \\
G35.59+0.06  & 18 56 04.01 & 02 23 26.1 & & G41.16--0.20 & 19 07 14.32 & 07 13 18.6 \\
G35.79--0.17 & 18 57 16.85 & 02 27 58.5 & & G41.23--0.20 & 19 07 21.50 & 07 17 08.6 \\
G36.02--0.20 & 18 57 45.85 & 02 39 06.1 & & G41.27+0.37  & 19 05 23.57 & 07 35 05.9 \\
G36.64--0.21 & 18 58 55.21 & 03 12 05.4 & & G41.34--0.14 & 19 07 21.78 & 07 25 17.9 \\
G36.70+0.09  & 18 57 59.09 & 03 24 06.7 & & G41.58+0.04  & 19 07 09.14 & 07 42 25.2 \\
G36.84--0.02 & 18 58 39.19 & 03 28 00.8 & & G42.03+0.19  & 19 07 28.15 & 08 10 54.6 \\
G36.92+0.48  & 18 56 59.77 & 03 46 04.9 & & G42.30--0.30 & 19 09 43.56 & 08 11 41.8 \\
G37.02--0.03 & 18 59 03.62 & 03 37 45.4 & & G42.70--0.15 & 19 09 55.12 & 08 36 54.0 \\
G37.04--0.04 & 18 59 04.37 & 03 38 33.2 & & G43.04--0.46 & 19 11 38.98 & 08 46 31.3 \\
G37.38--0.09 & 18 59 51.55 & 03 55 18.9 & & G43.08--0.08 & 19 10 22.02 & 08 58 51.9 \\
G37.47--0.11 & 19 00 07.10 & 03 59 53.7 & & G44.31+0.04  & 19 12 15.67 & 10 07 54.1 \\
G37.60+0.42  & 18 58 26.76 & 04 20 46.0 & & G44.64--0.52 & 19 14 53.73 & 10 10 08.2 \\
G37.74--0.12 & 19 00 36.88 & 04 13 20.9 & & G45.47+0.05  & 19 14 24.11 & 11 09 43.8 \\
G37.76--0.19 & 19 00 55.38 & 04 12 13.6 & & G45.49+0.13  & 19 14 11.32 & 11 13 07.0 \\
G38.03--0.30 & 19 01 50.43 & 04 24 19.6 & & G45.57--0.12 & 19 15 13.12 & 11 10 17.6 \\
G38.08--0.27 & 19 01 47.28 & 04 27 21.6 & & G45.81--0.36 & 19 16 31.13 & 11 16 11.3 \\
G38.12--0.24 & 19 01 44.12 & 04 30 38.1 & & G46.07+0.22  & 19 14 56.03 & 11 46 13.9 \\
G38.20--0.08 & 19 01 18.7  & 04 39 37   & & G46.12+0.38  & 19 14 25.52 & 11 53 27.8 \\
G38.26--0.08 & 19 01 26.21 & 04 42 18.0 & & G48.90--0.27 & 19 22 10.27 & 14 02 43.9 \\
G38.26--0.20 & 19 01 52.92 & 04 38 40.2 & & G49.27+0.31  & 19 20 44.83 & 14 38 27.2 \\
G38.56+0.15  & 19 01 08.39 & 05 04 37.8 & & G49.35+0.41  & 19 20 32.41 & 14 45 46.1 \\
G38.60--0.21 & 19 02 33.43 & 04 56 37.0 & & G49.42+0.32  & 19 20 59.78 & 14 46 49.8 \\
G38.66+0.08  & 19 01 35.23 & 05 07 47.8 & & G49.60--0.25 & 19 23 26.64 & 14 40 16.9 \\
G38.92--0.36 & 19 03 38.59 & 05 09 42.3 & & G49.62--0.36 & 19 23 52.76 & 14 38 03.7 \\
G39.39--0.14 & 19 03 45.22 & 05 40 43.6 & & G50.78+0.15  & 19 24 17.35 & 15 54 02.2 \\
G40.28--0.22 & 19 05 41.16 & 06 26 13.8 & & G52.92+0.41  & 19 27 34.91 & 17 54 38.5 \\
G40.94--0.04 & 19 06 15.33 & 07 05 54.8 & & G53.04+0.11  & 19 28 55.59 & 17 52 01.8 \\
G41.08--0.13 & 19 06 49.02 & 07 11 07.0 & & G53.62+0.04  & 19 30 22.99 & 18 20 27.5 \\
\enddata
\tablecomments{The columns show the source name 6.7~GHz methanol maser and the equatorial coordinates of the MIPS 24~$\mu$m point source counterpart. Only sources for which a reliable counterpart could be identified are shown here. Sources that are completely saturated and blanked, or which lie in regions of extended emission are omitted.}
\end{deluxetable}

\begin{deluxetable}{ccc}
\tabletypesize{\footnotesize}
\tablecaption{Suggested coordinates for 6.7~GHz methanol masers that were not detected using MERLIN and EVLA.\label{otherpos}}
\tablewidth{0pt}
\tablehead{
\colhead{Source name} & \colhead{RA (J2000)} & \colhead{Dec. (J2000)} \\
\colhead{} & \colhead{(h m s)} & \colhead{(\degr~\arcmin~\arcsec)} 
}
\startdata
G34.82+0.35  & 18 53 37.9 & 01 50 31 \\
G35.39+0.02  & 18 55 50.7 & 02 11 40 \\
G36.90--0.41\tablenotemark{a} & 19 00 08.4 & 03 20 33 \\
G39.54--0.38 & 19 04 52.6 & 05 42 08 \\
G41.87--0.10 & 19 08 10.4 & 07 54 00 \\
G48.89--0.17 & 19 21 47.5 & 14 05 03 \\
\enddata
\tablecomments{Based on the properties of 24~$\mu$m counterparts of 6.7~GHz methanol masers, the positions quoted here are expected to be accurate to $\sim 1\arcsec$.}
\tablenotetext{a}{A second 24~$\mu$m point source is located within 18$''$ of the single dish position at $19^h00^m07^s.9$, $03\degr 20\arcmin 36\arcsec$; see Sect. \ref{mirassociation}.}
\end{deluxetable}

\begin{figure}
\centering
\includegraphics[height=0.8\textheight]{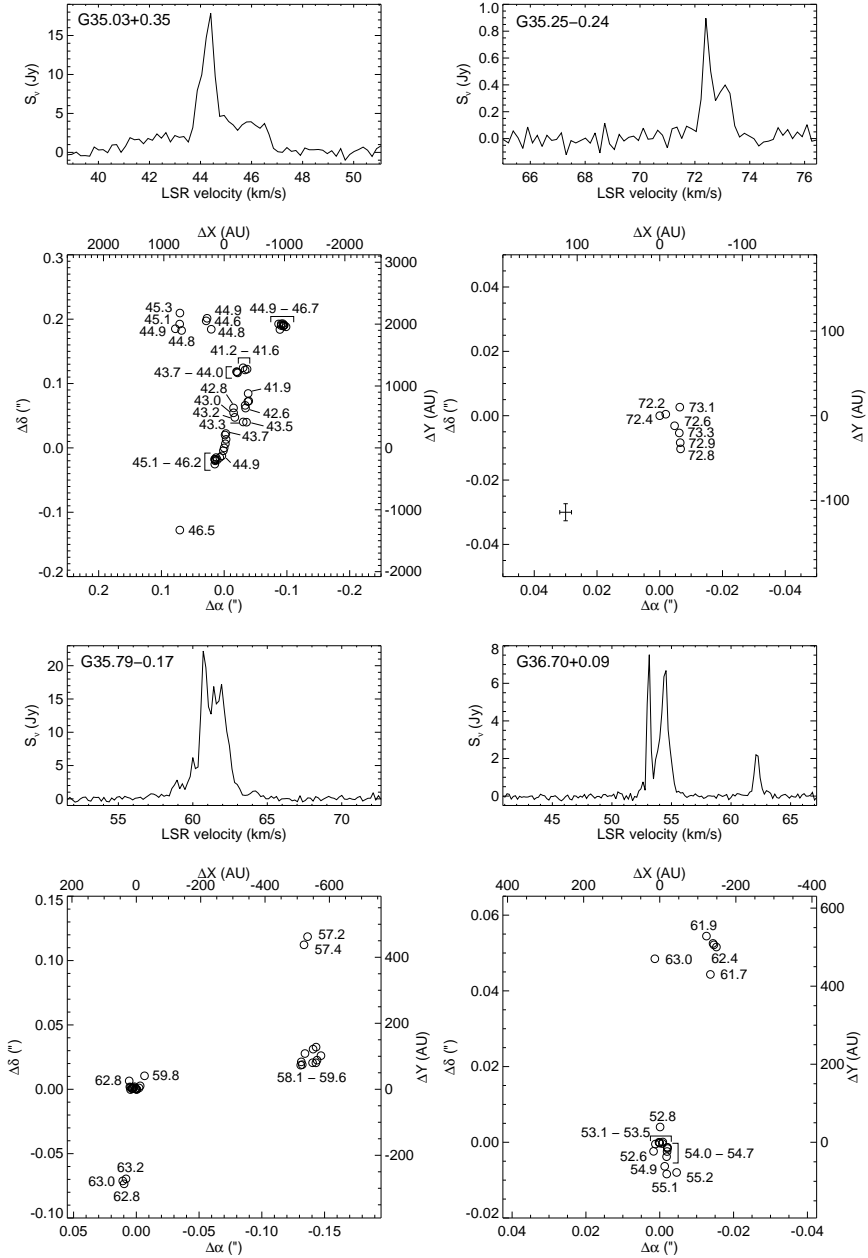}
\caption{Morphologies of maser spots as seen by MERLIN. For each source, the upper panel shows the net spectrum, while the lower panel shows the locations of the maser spots. Typical uncertainties in the spot positions are shown in the lower left when comparable to the spot symbols. The velocities of the spots are indicated adjacent to the spots themselves. When there is a linear velocity trend across a set of spots, only the initial and final velocities are indicated at the ends of the feature. Only sources which have at least 4 maser spots detected at the 6$\sigma$ level are included in this figure.}\label{merlinspotmaps}
\end{figure}

\setcounter{figure}{0}
\begin{figure}
\centering
\includegraphics[height=0.95\textheight]{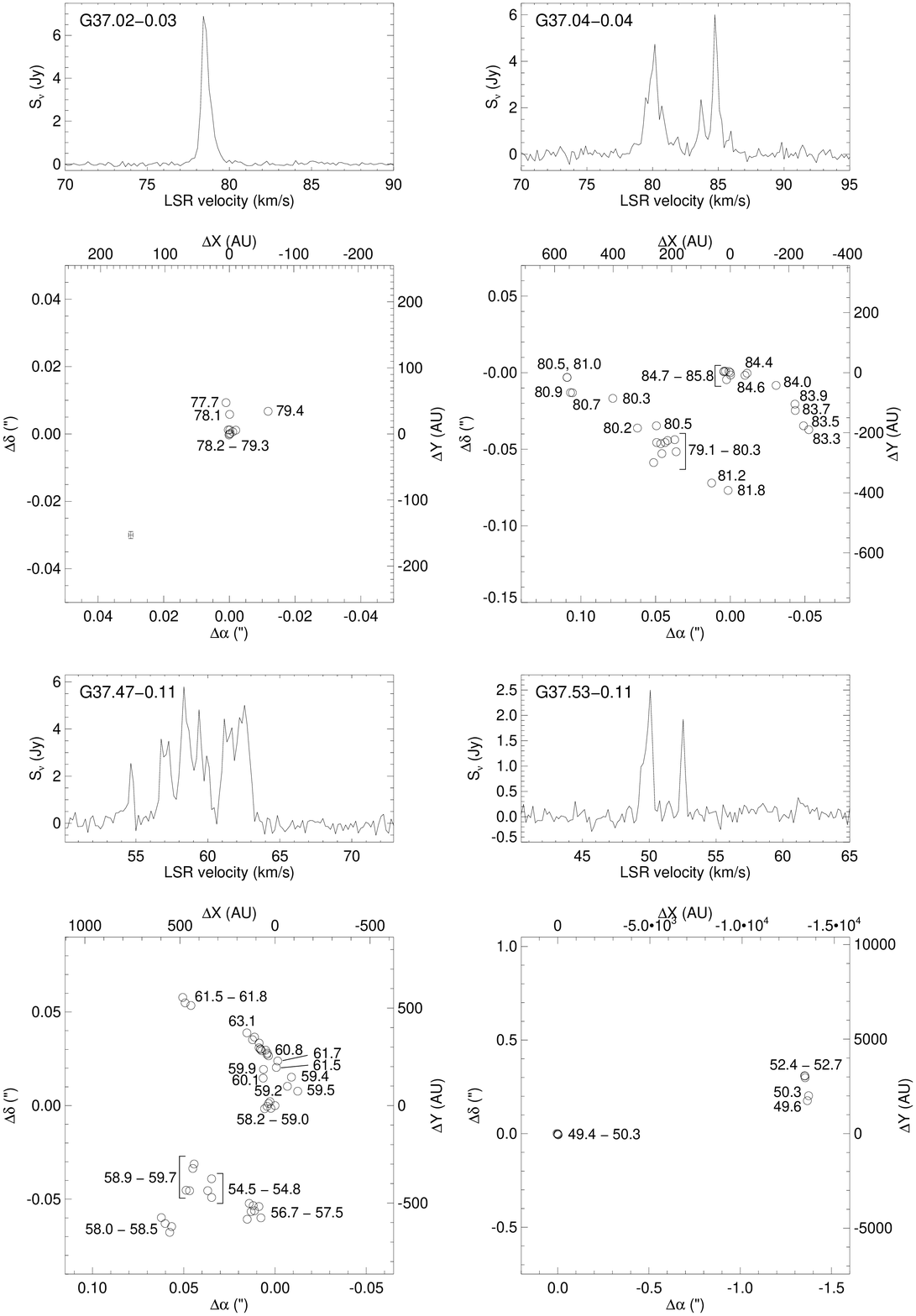}
\caption{Contd.}
\end{figure}

\setcounter{figure}{0}
\begin{figure}
\centering
\includegraphics[height=0.95\textheight]{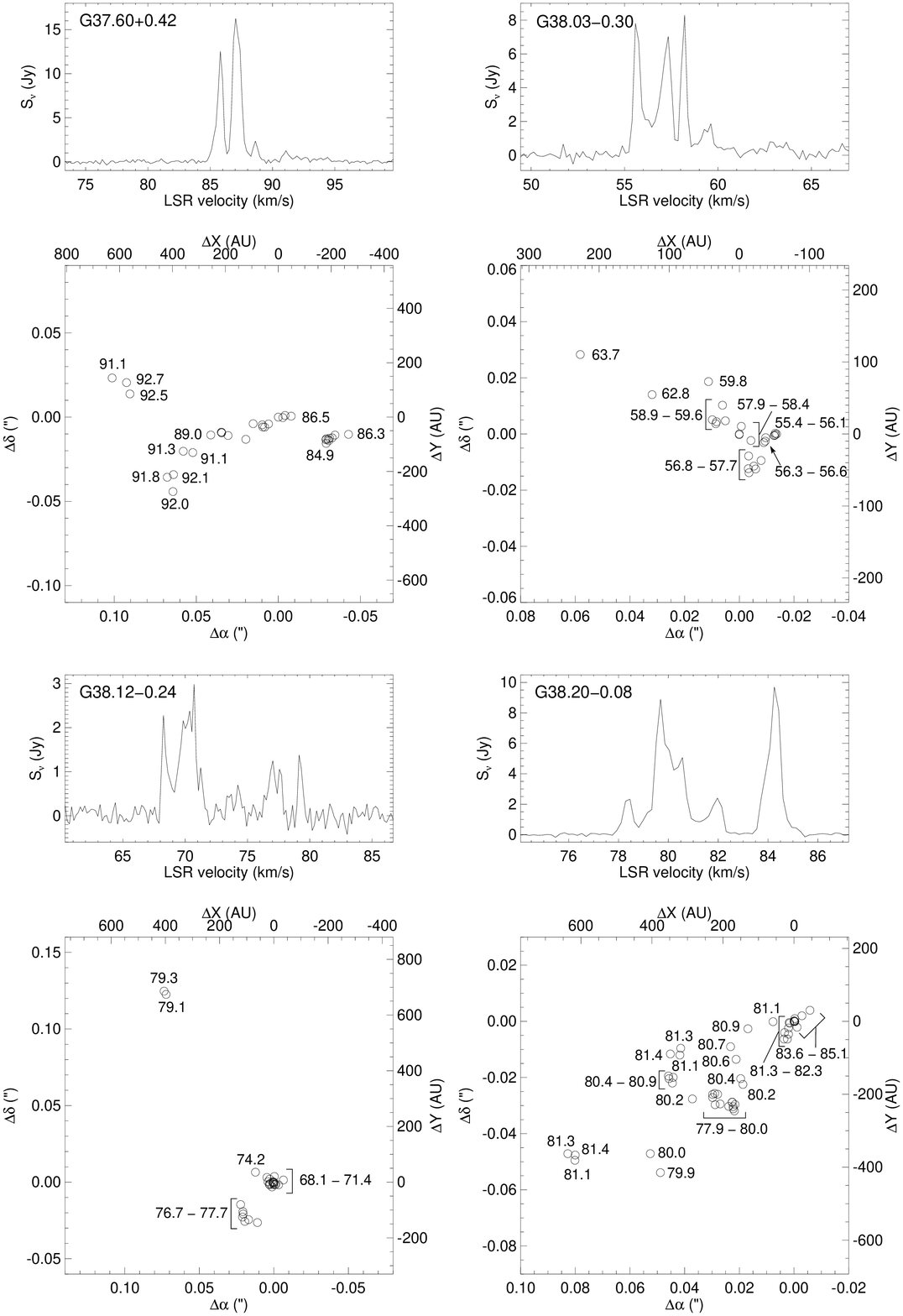}
\caption{Contd.}
\end{figure}

\setcounter{figure}{0}
\begin{figure}
\centering
\includegraphics[height=0.95\textheight]{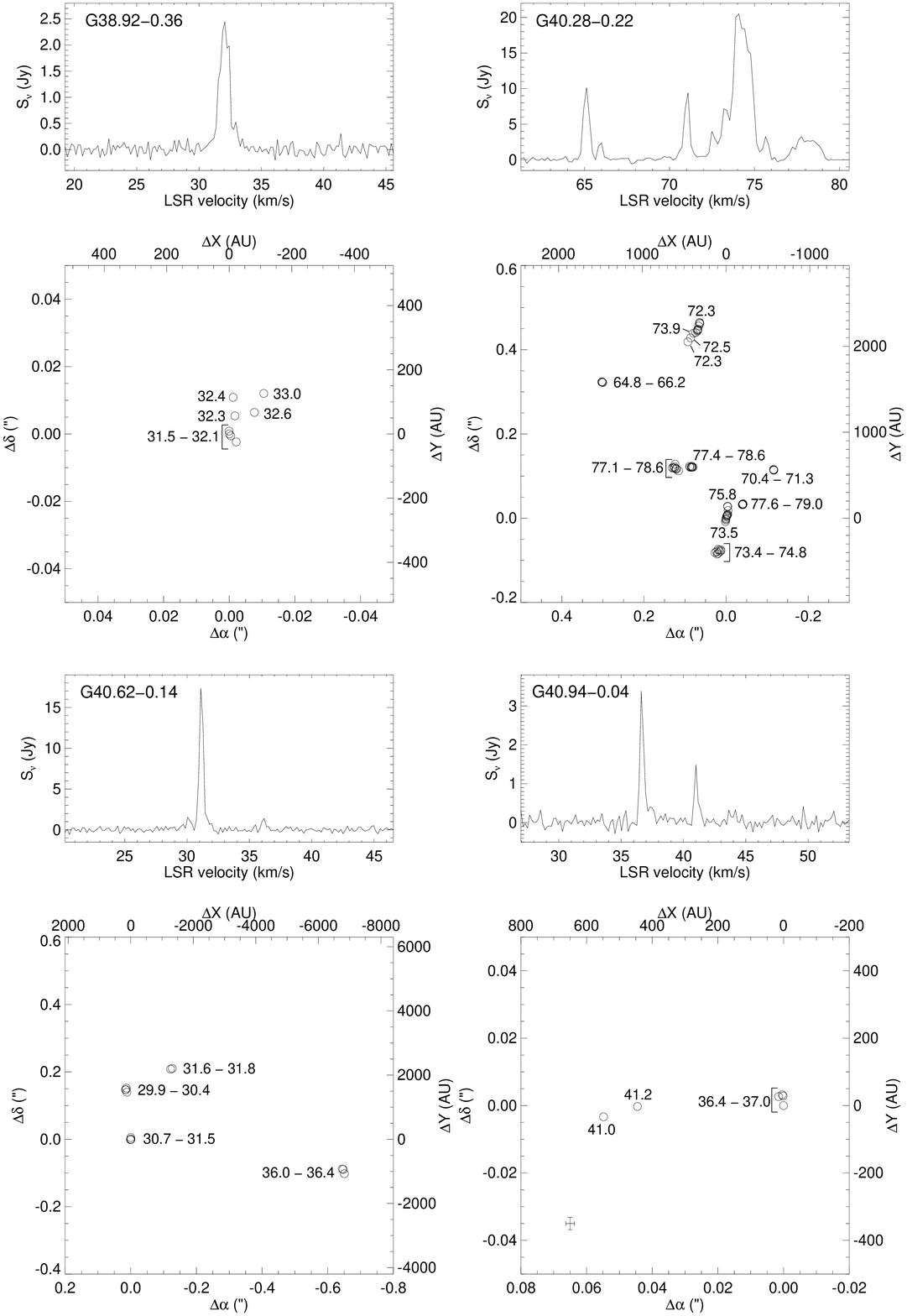}
\caption{Contd.}
\end{figure}

\setcounter{figure}{0}
\begin{figure}
\centering
\includegraphics[height=0.95\textheight]{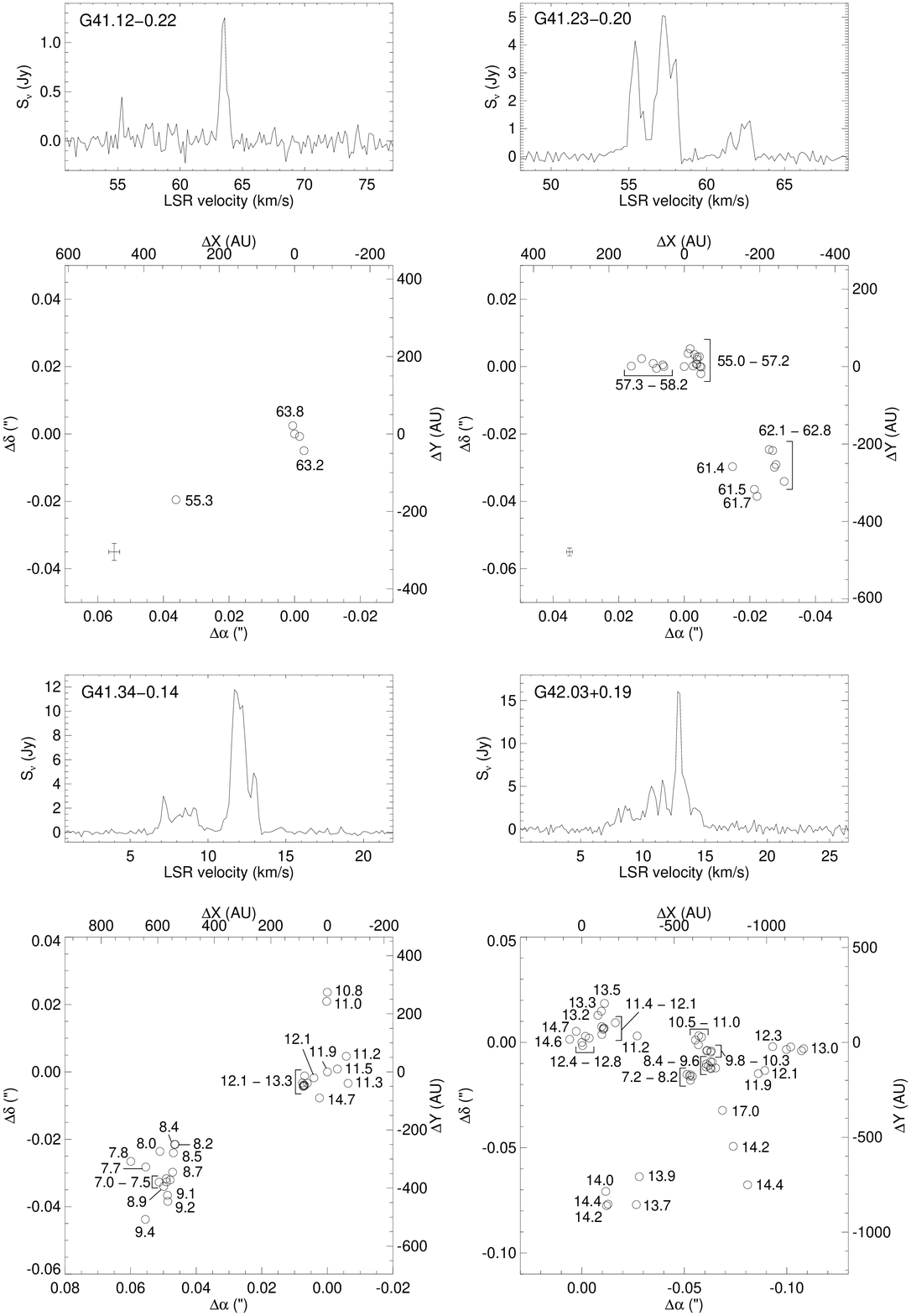}
\caption{Contd.}
\end{figure}

\setcounter{figure}{0}
\begin{figure}
\centering
\includegraphics[height=0.95\textheight]{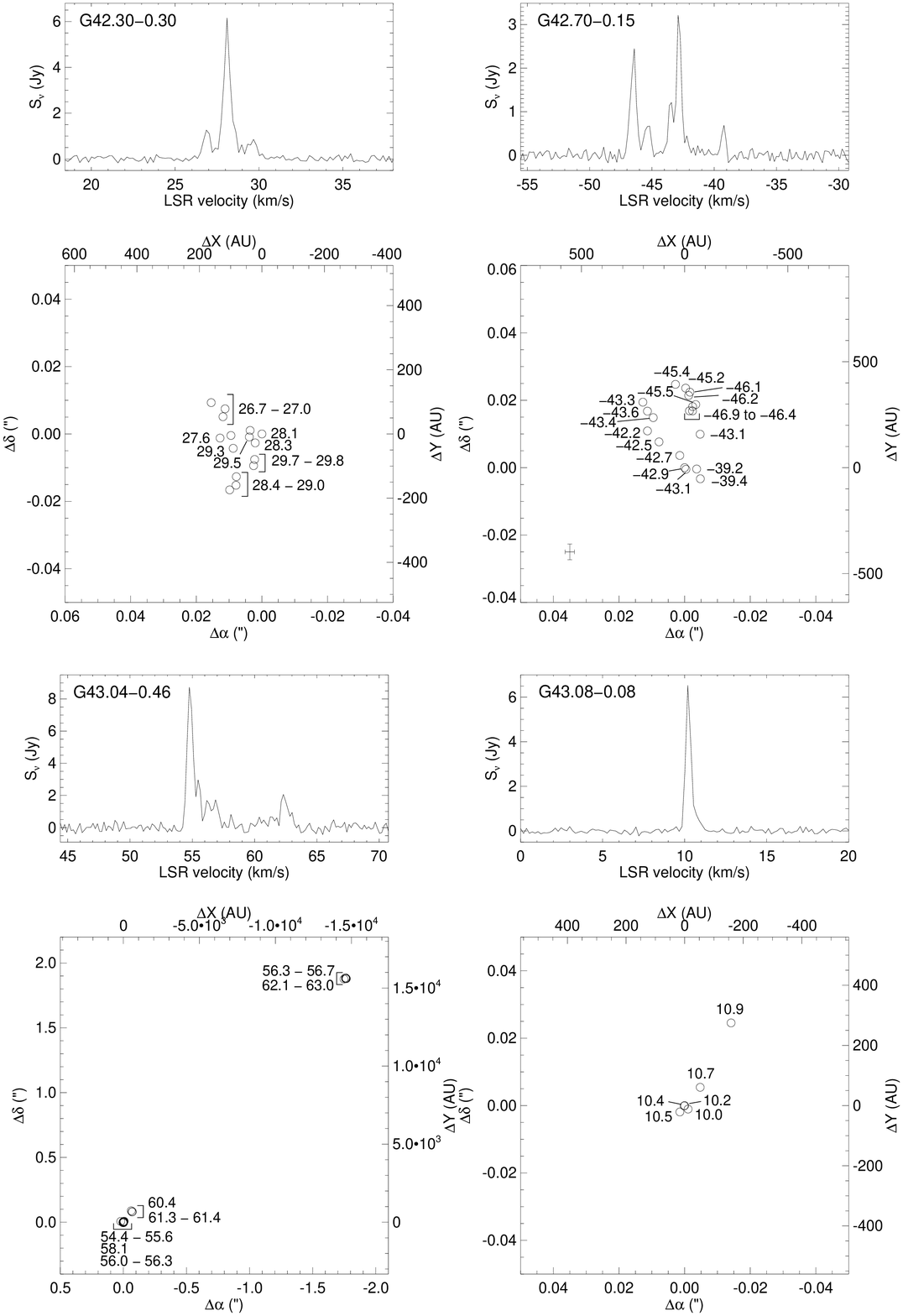}
\caption{Contd.}
\end{figure}

\setcounter{figure}{0}
\begin{figure}
\centering
\includegraphics[height=0.95\textheight]{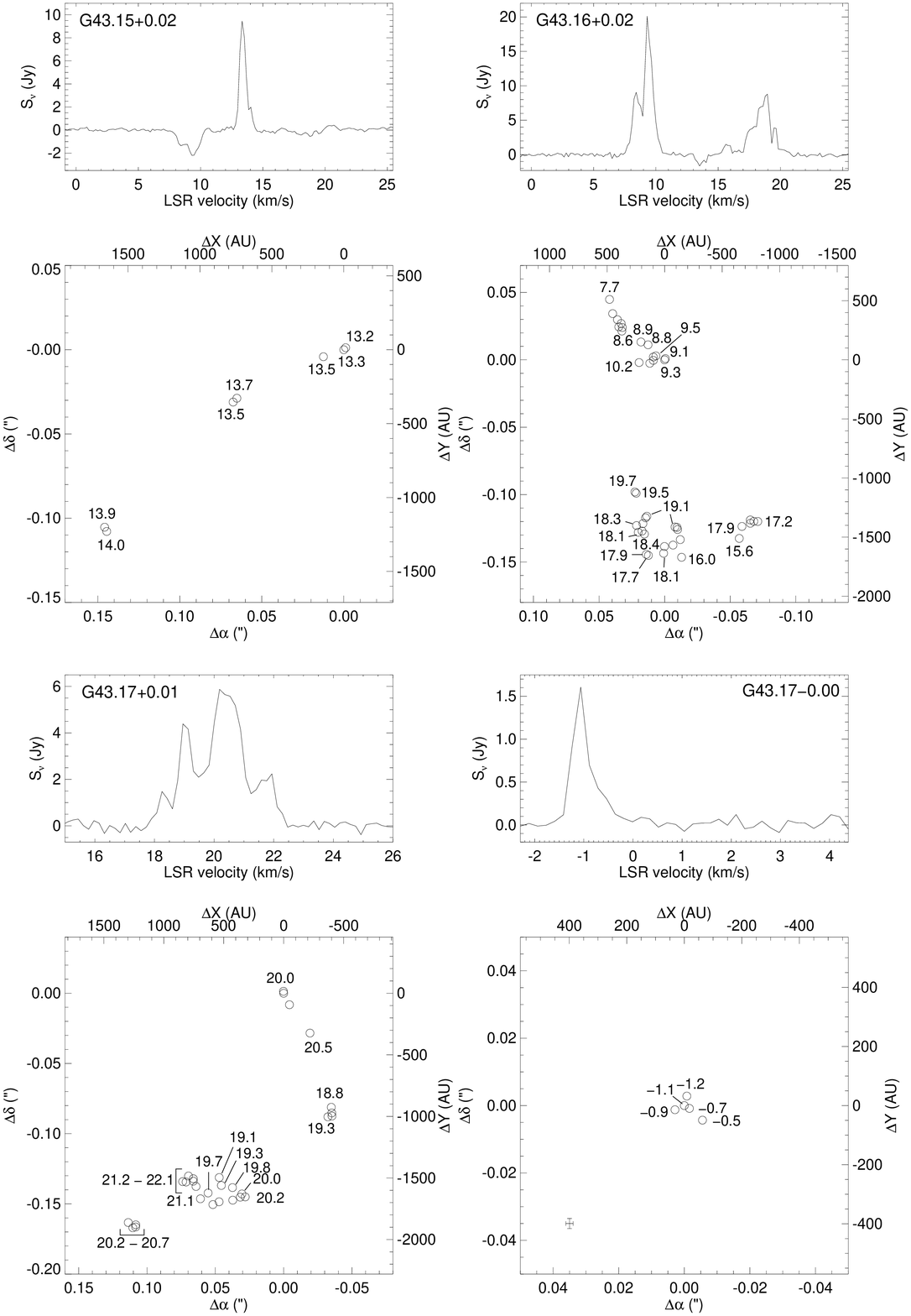}
\caption{Contd.}
\end{figure}

\setcounter{figure}{0}
\begin{figure}
\centering
\includegraphics[height=0.95\textheight]{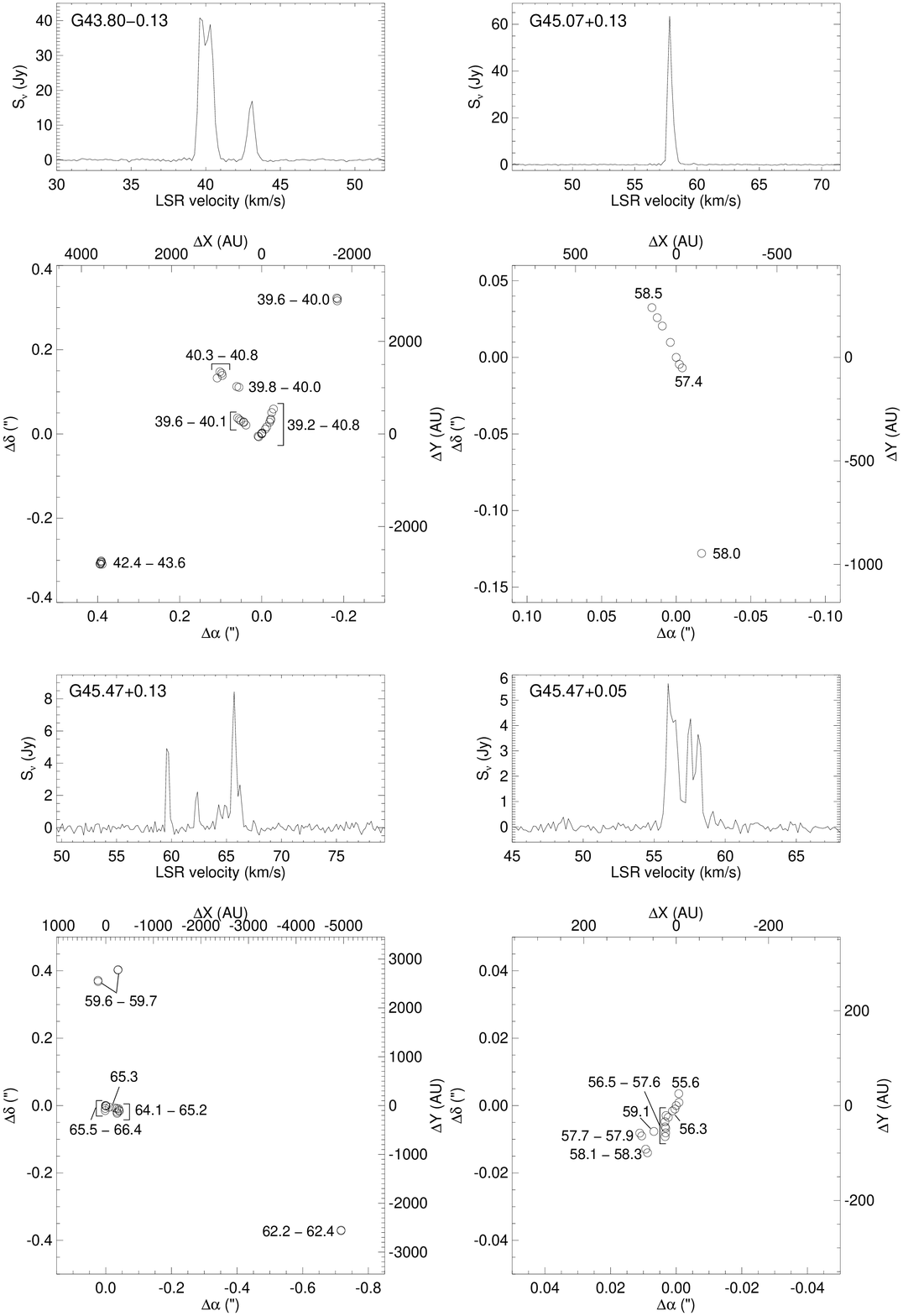}
\caption{Contd.}
\end{figure}

\setcounter{figure}{0}
\begin{figure}
\centering
\includegraphics[height=0.95\textheight]{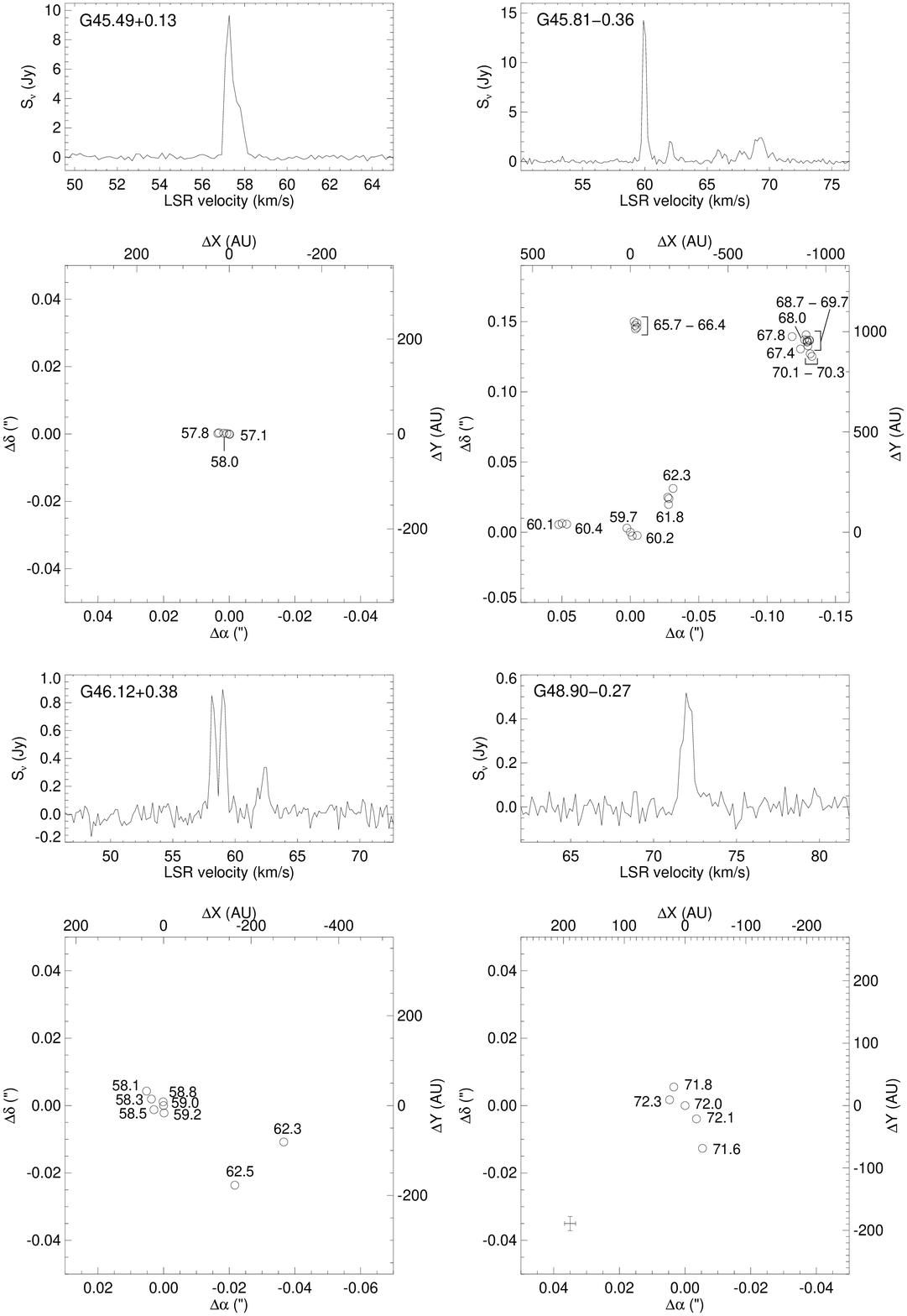}
\caption{Contd.}
\end{figure}

\setcounter{figure}{0}
\begin{figure}
\centering
\includegraphics[height=0.95\textheight]{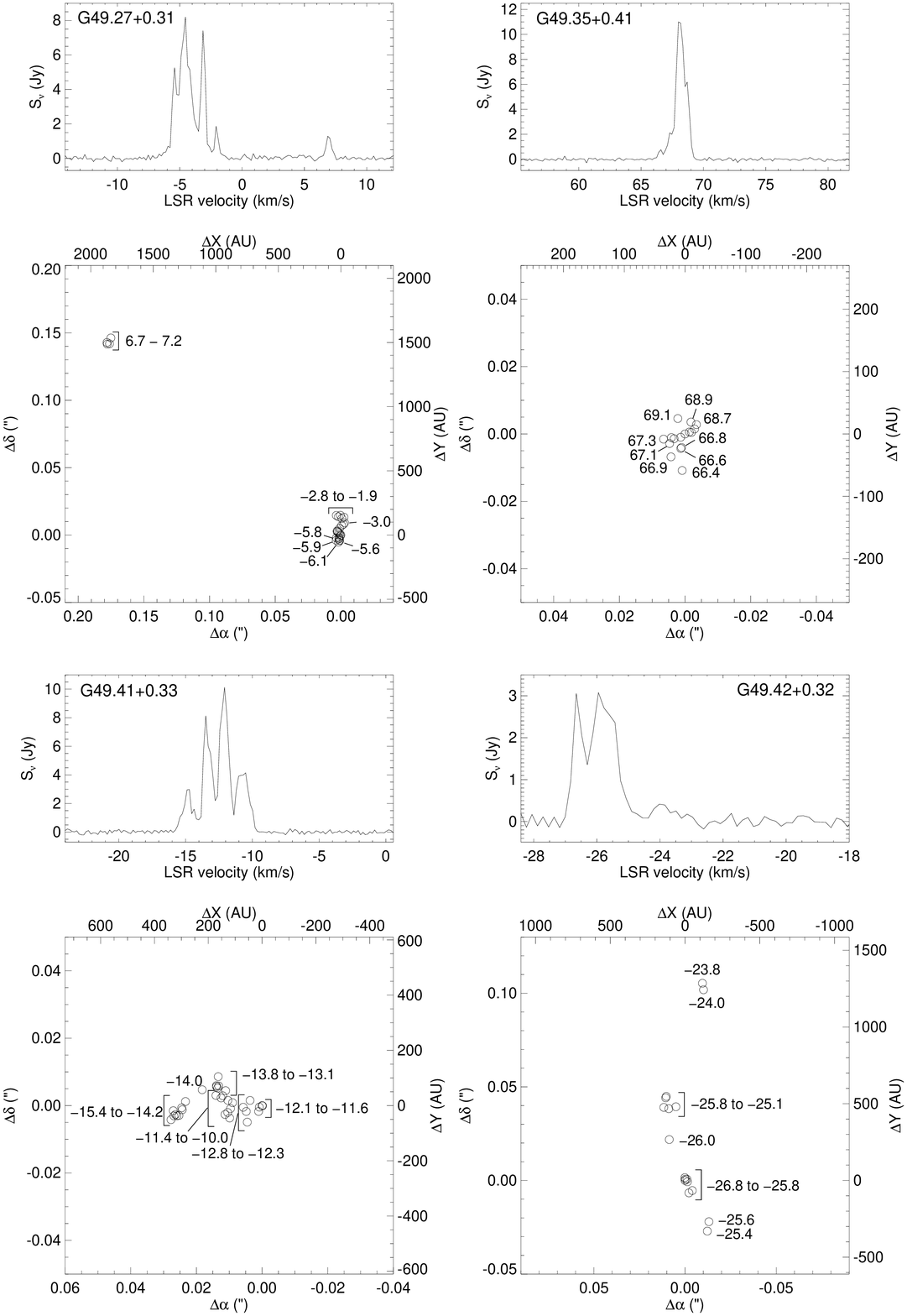}
\caption{Contd.}
\end{figure}

\setcounter{figure}{0}
\begin{figure}
\centering
\includegraphics[height=0.95\textheight]{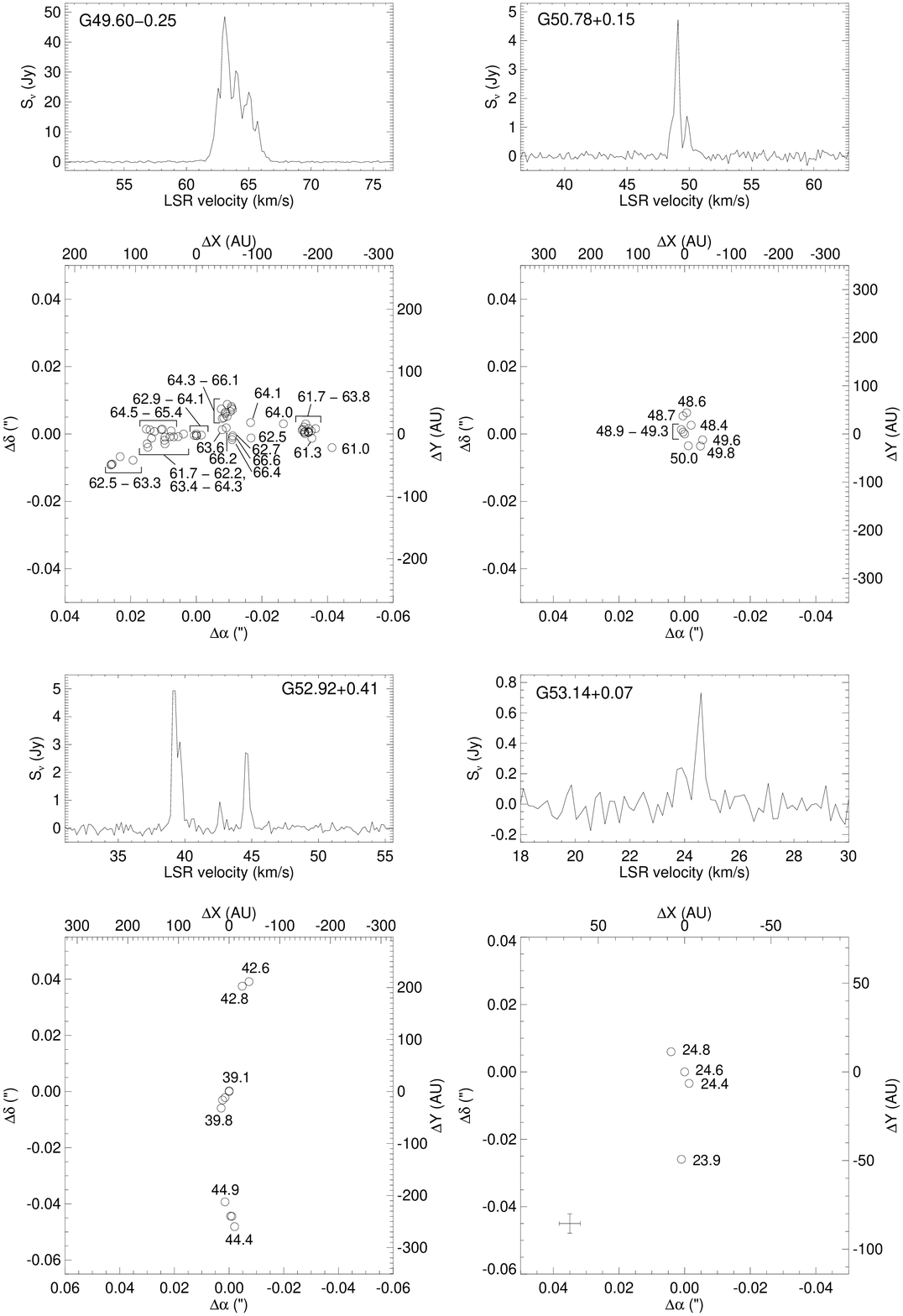}
\caption{Contd.}
\end{figure}

\setcounter{figure}{0}
\begin{figure}
\centering
\includegraphics[height=0.47\textheight]{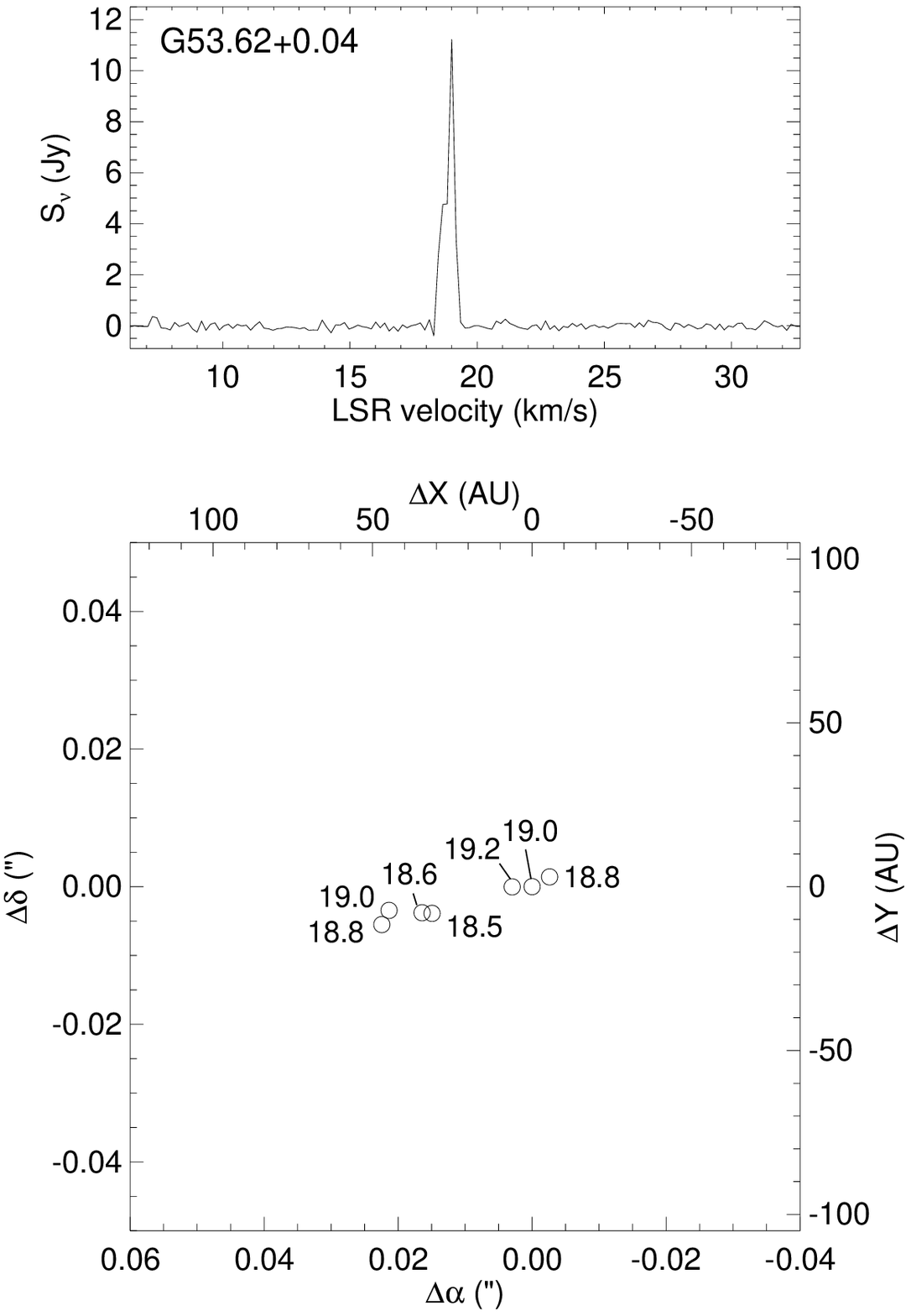}
\caption{Contd.}
\end{figure}
\clearpage

\begin{figure}
\centering
\includegraphics[height=0.85\textheight]{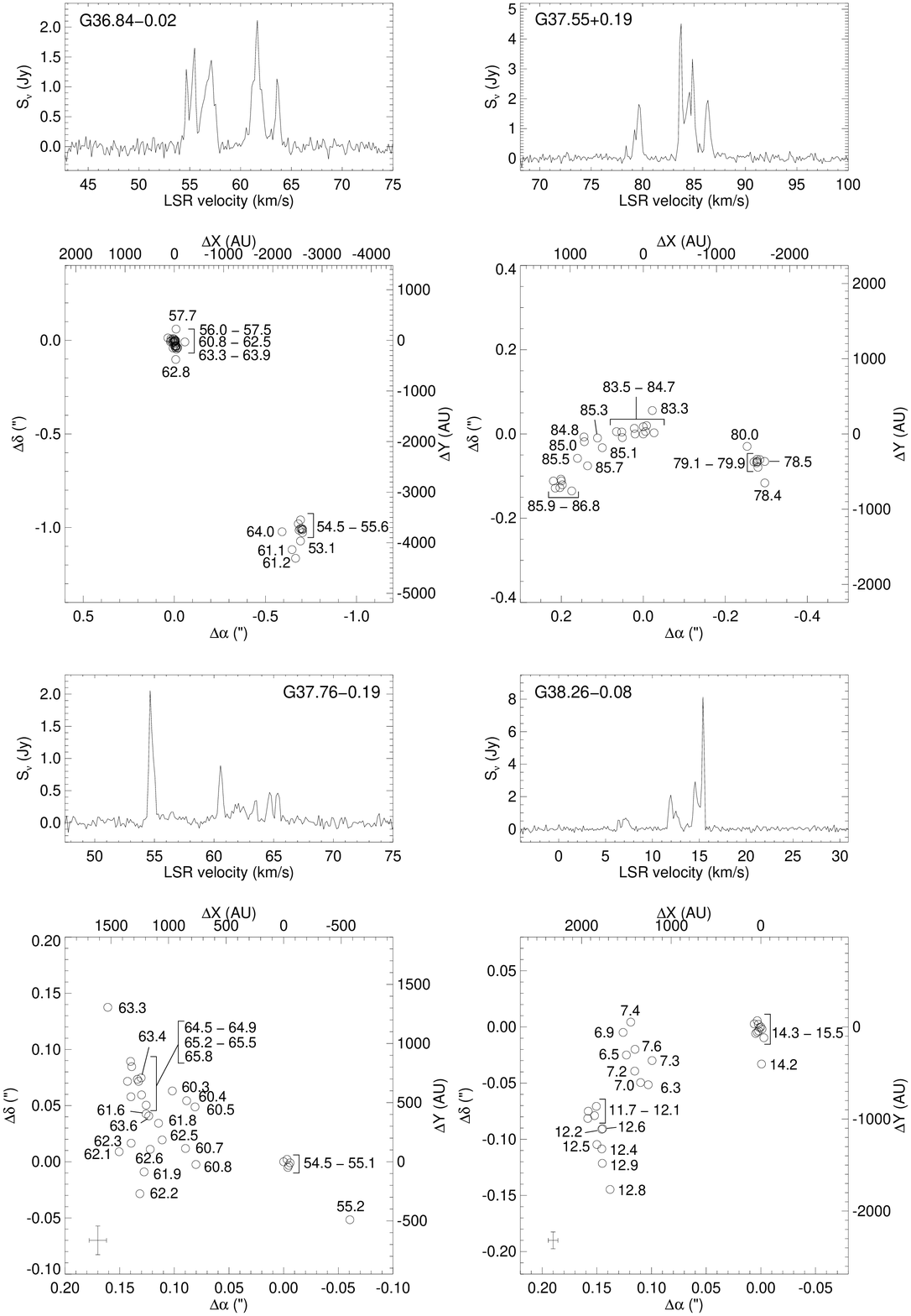}
\caption{Morphologies of maser spots as seen by EVLA. As in Fig.~\ref{merlinspotmaps}, the upper panel shows the net spectrum of each source, while the lower panel shows the locations of the maser spots. Typical uncertainties in the spot positions are shown in the lower left when comparable to the spot symbols. The velocities of the spots are indicated adjacent to the spots themselves.}\label{evlaspotmaps}
\end{figure}

\setcounter{figure}{1}
\begin{figure}
\centering
\includegraphics[height=0.47\textheight]{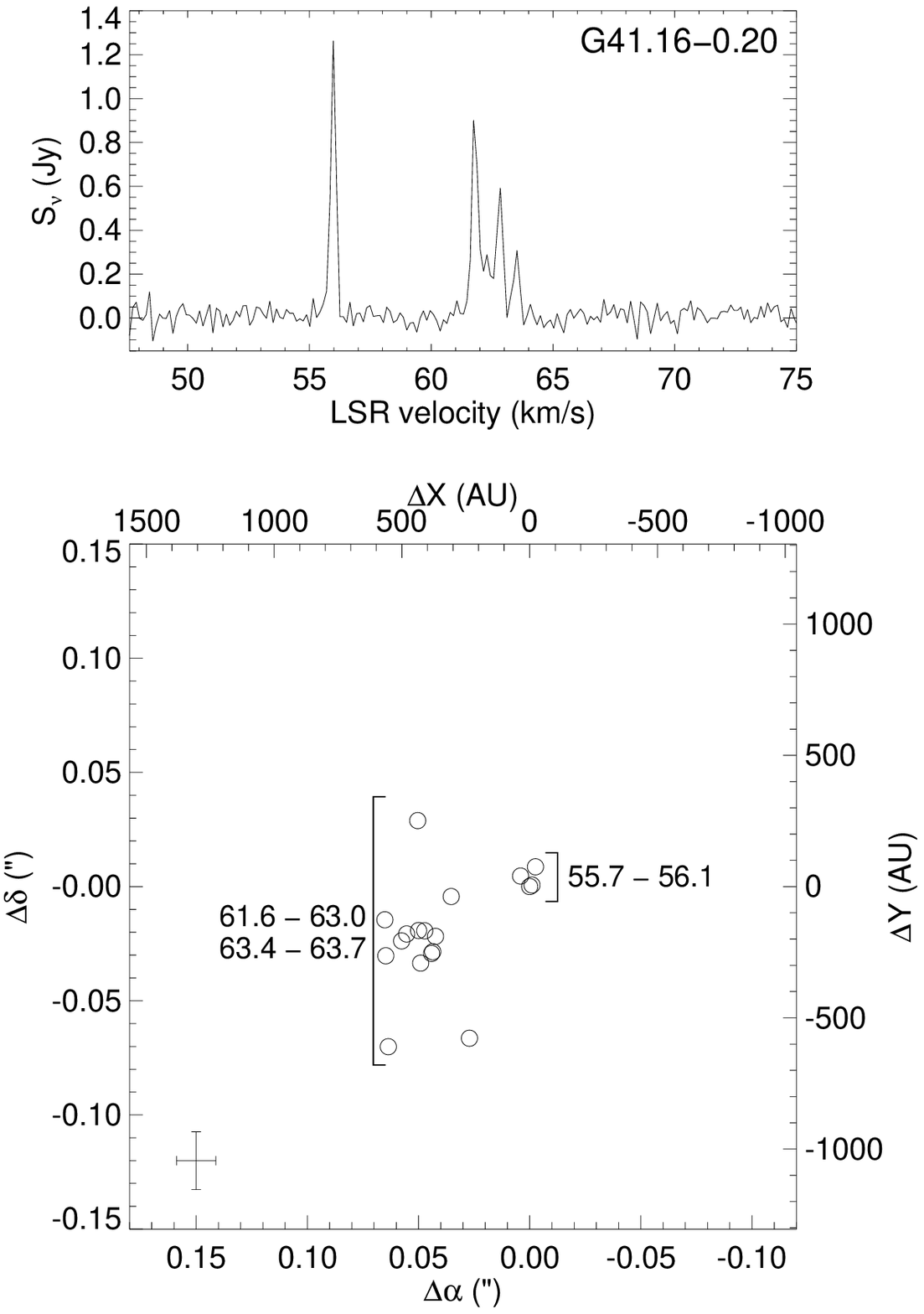}
\caption{Contd.}
\end{figure}
\clearpage

\begin{figure}
\centering
\includegraphics[width=\textwidth]{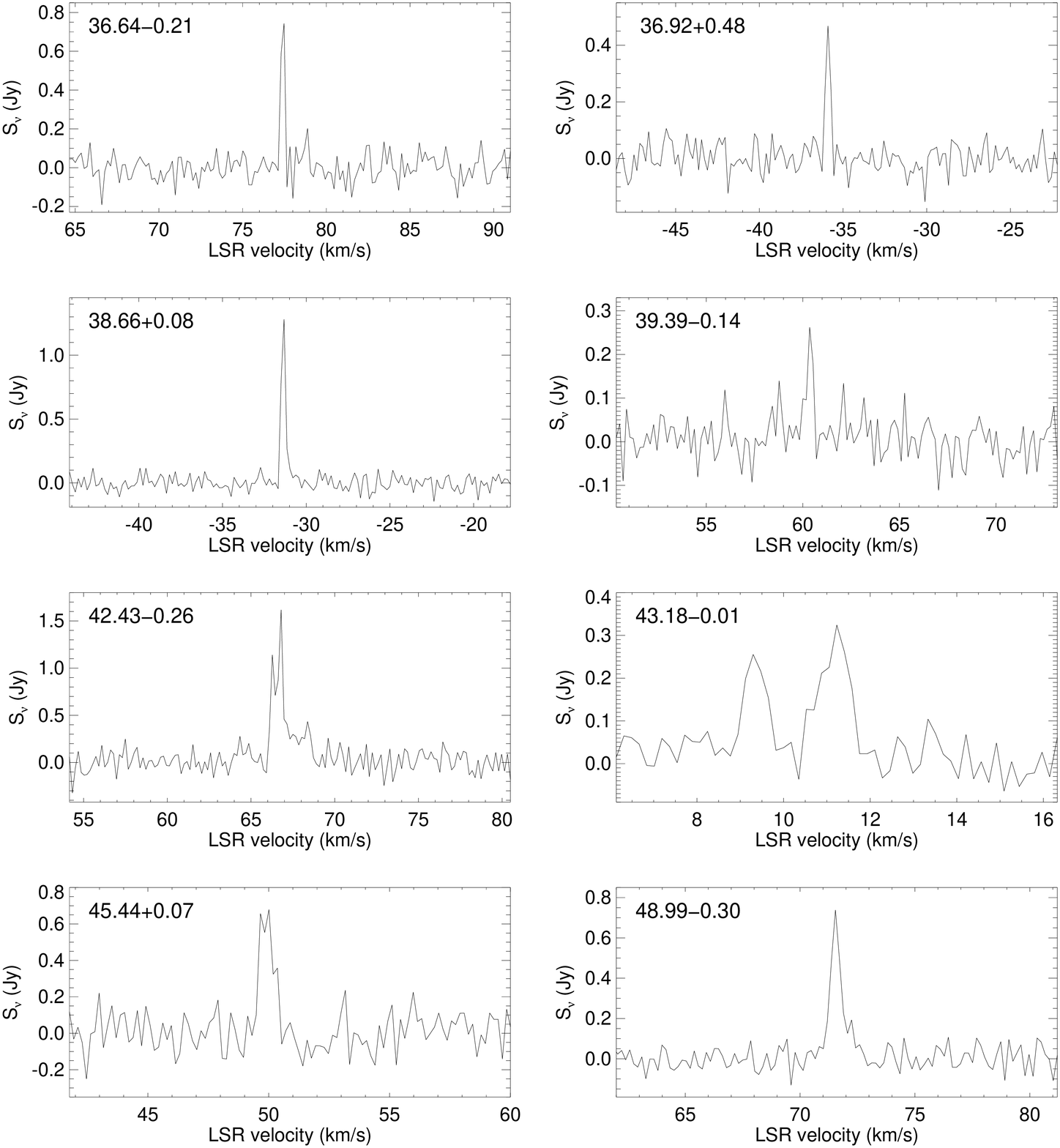}
\caption{Spectra of sources detected in the MERLIN data that have fewer than 4 maser spots detected at the $6\sigma$ level.}\label{merlinspectra}
\end{figure}
\clearpage

\begin{figure}
\centering
\includegraphics[width=\textwidth]{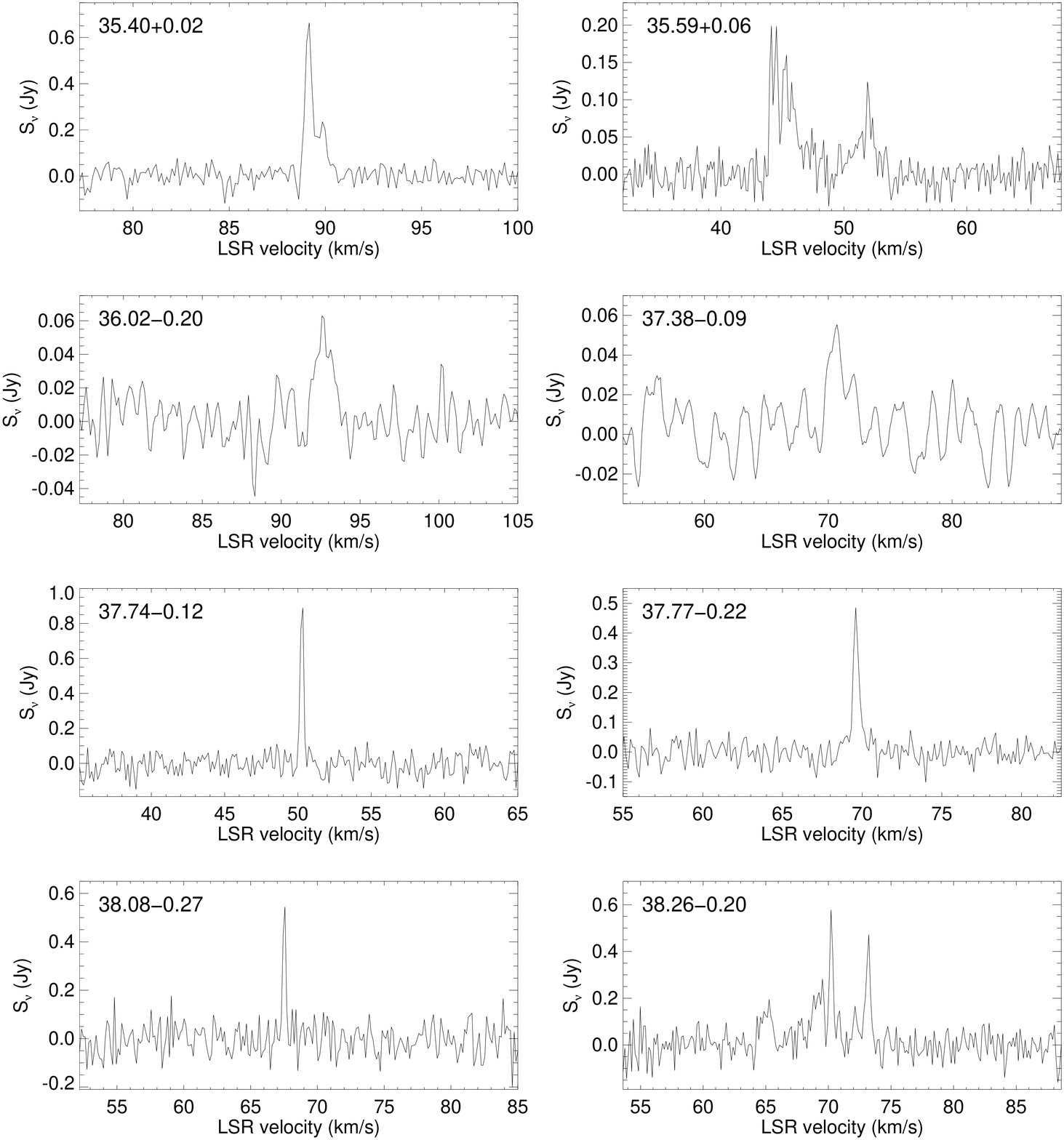}
\caption{Spectra of sources detected in the EVLA data for which maser spot morphologies cannot be reliably determined due to constraints of angular resolution and/or signal to noise ratio.}\label{evlaspectra}
\end{figure}

\setcounter{figure}{3}
\begin{figure}
\centering
\includegraphics[width=\textwidth]{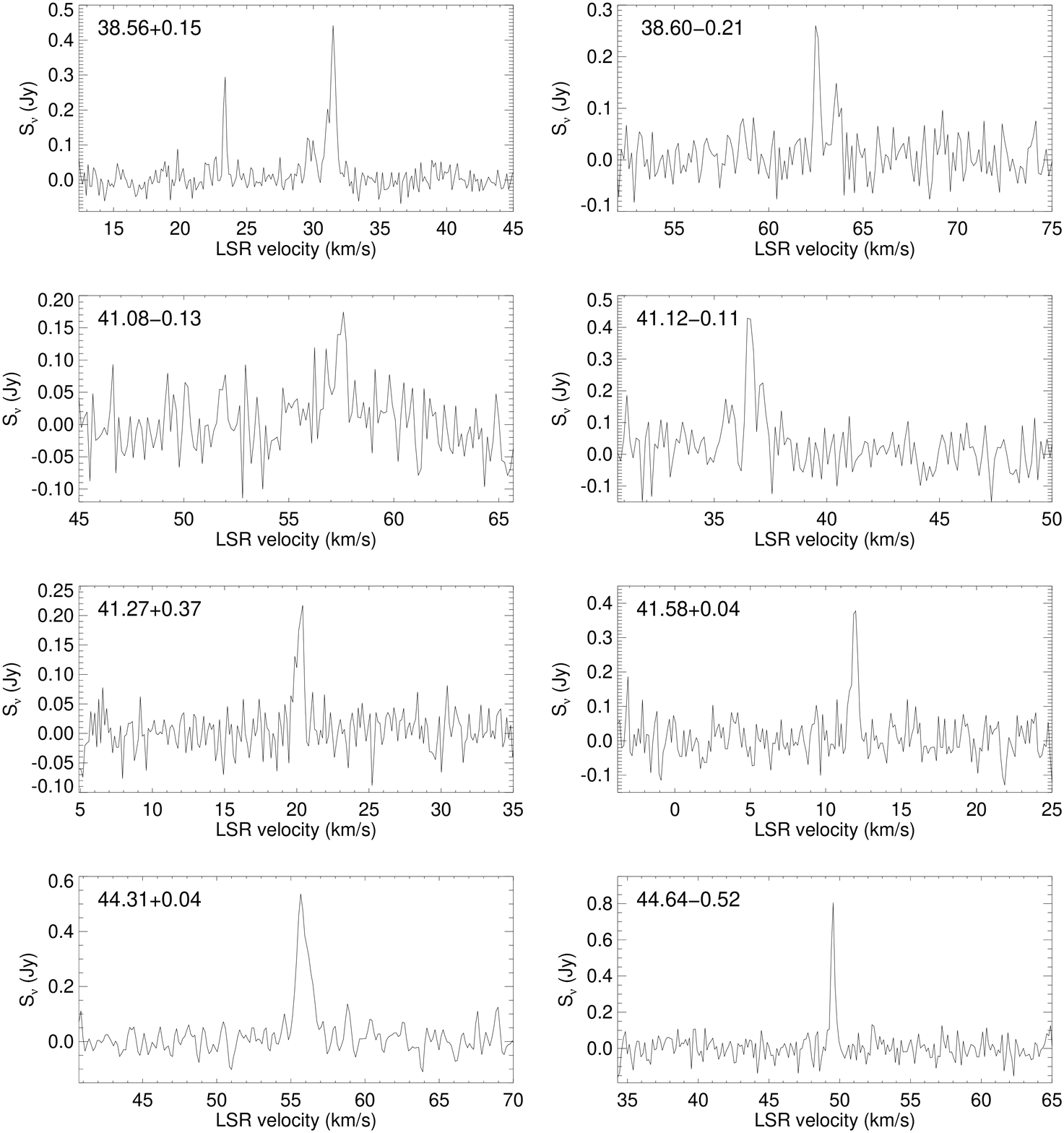}
\caption{Contd.}
\end{figure}

\setcounter{figure}{3}
\begin{figure}
\centering
\includegraphics[width=\textwidth]{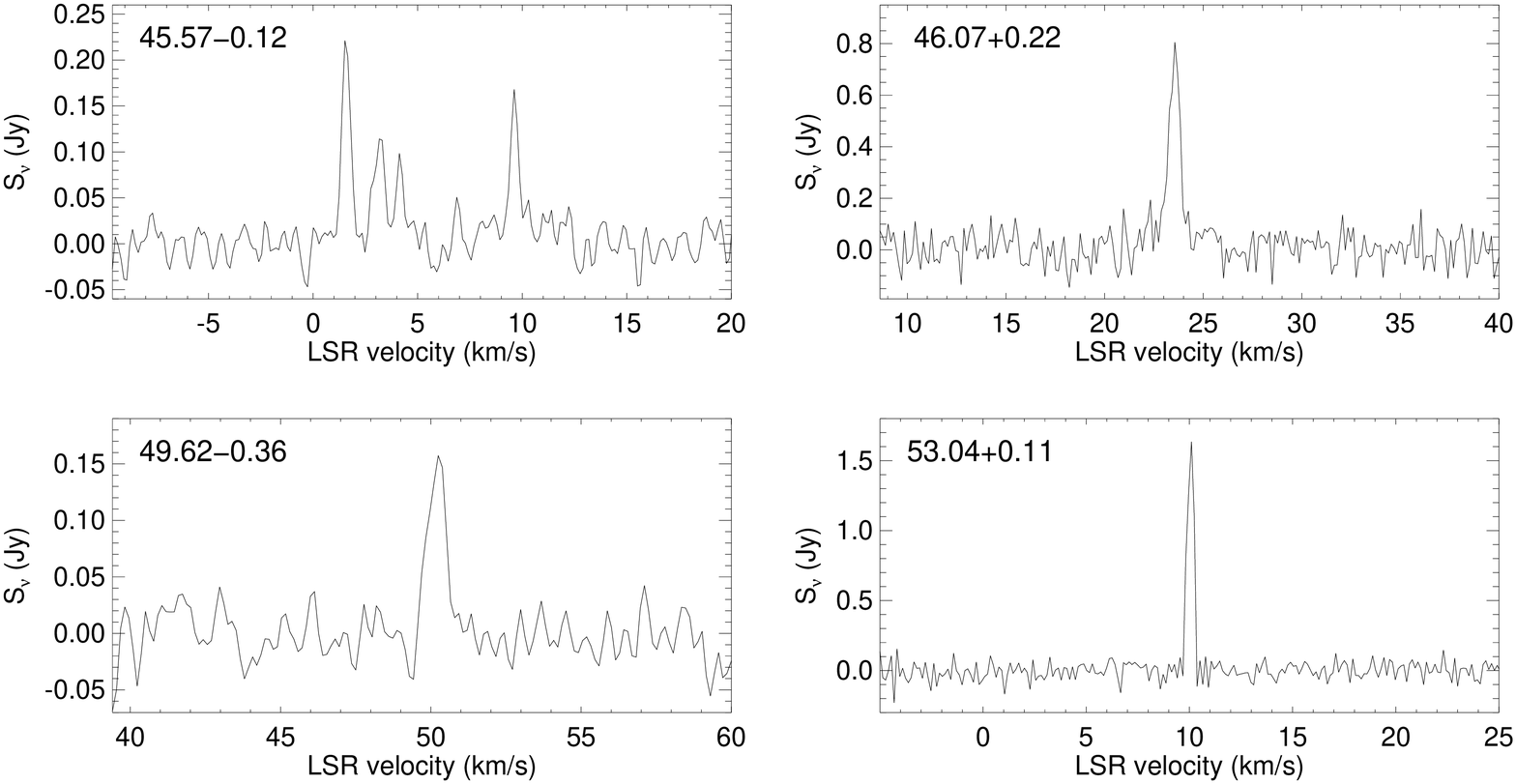}
\caption{Contd.}
\end{figure}
\clearpage

\begin{figure}
\centering
\includegraphics[width=0.45\textwidth]{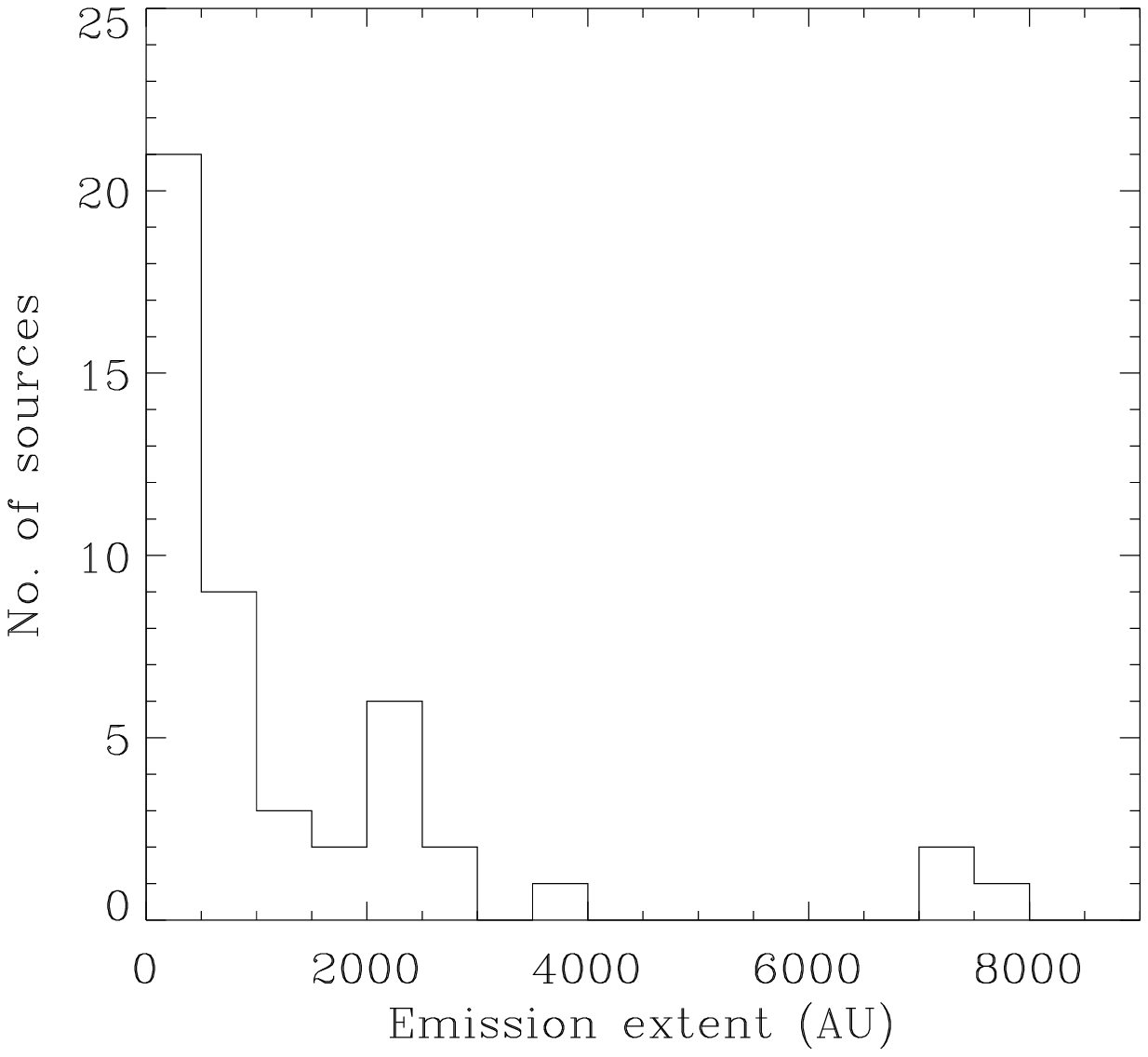}
\caption{Histogram of the spatial extent of maser emission along the major axis of the spot distribution.}\label{majexthist}
\end{figure}

\begin{figure}
\centering
\includegraphics[width=0.45\textwidth]{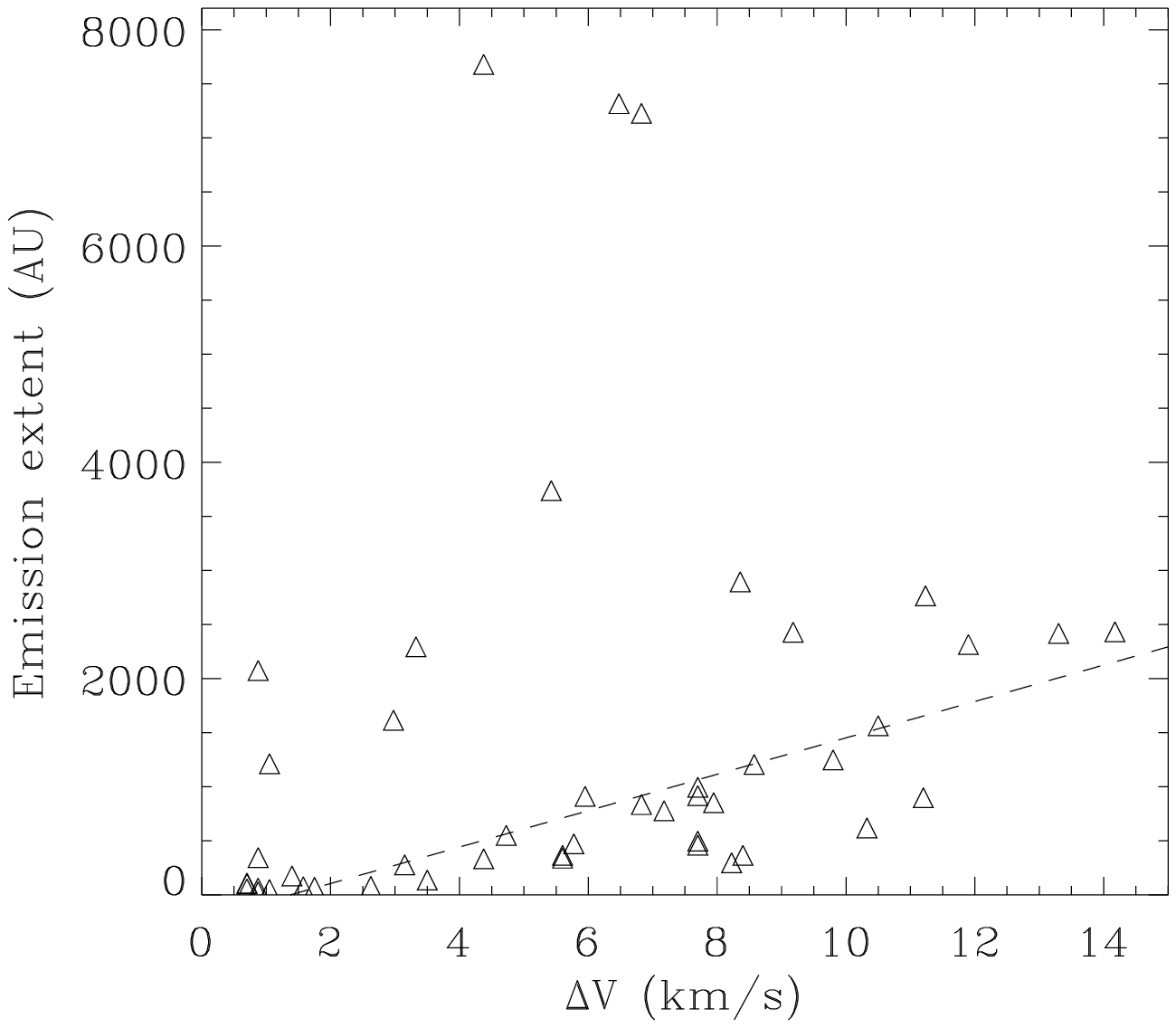}
\caption{The spatial major axis extent of maser emission as a function of the velocity width of the maser, defined from the minimum and maximum velocities of maser emission. The dashed line shows a linear fit to the data after excluding outliers.}\label{sizevelcorr}
\end{figure}

\begin{figure}
\centering
\includegraphics[width=0.45\textwidth]{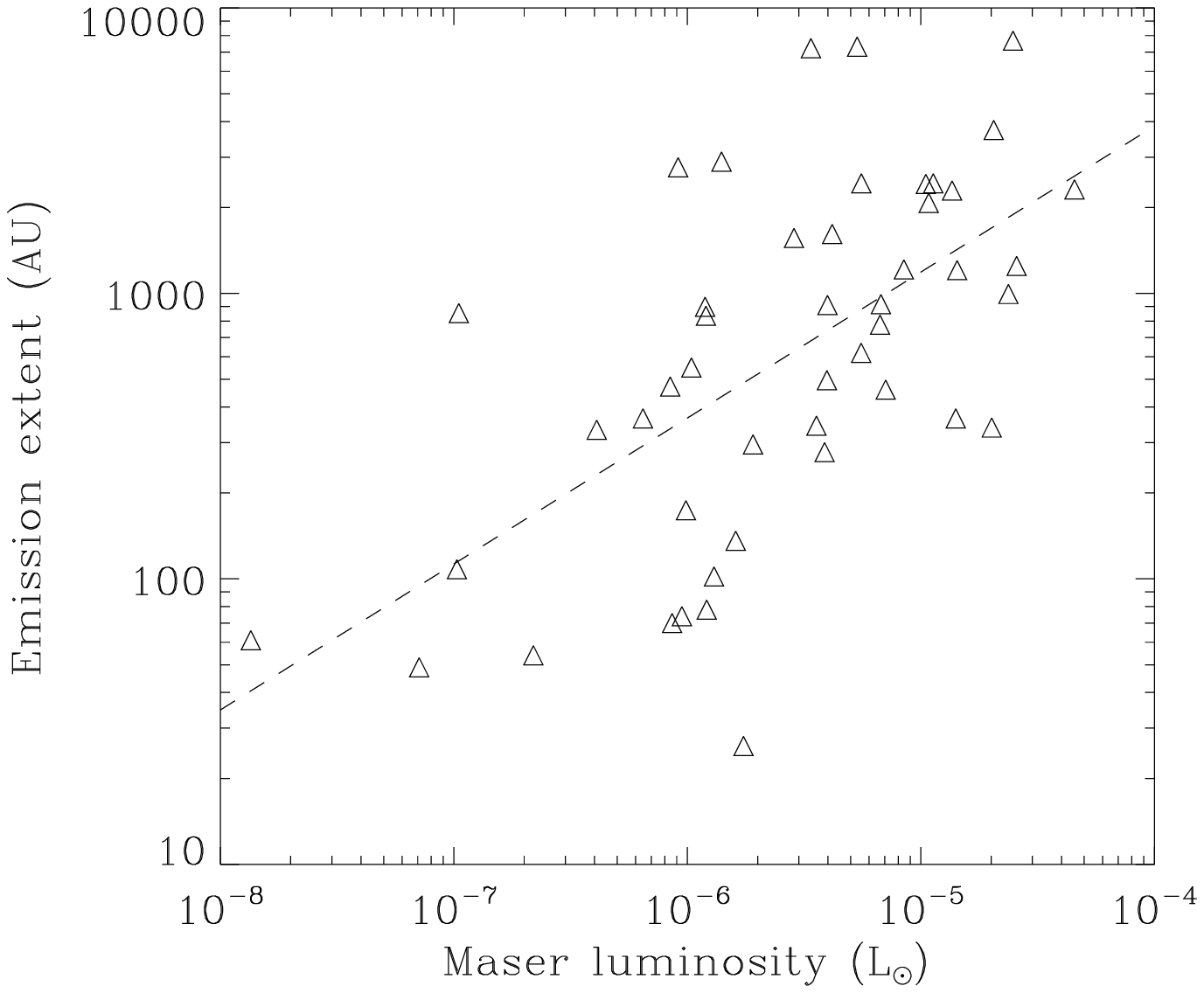}
\caption{The spatial major axis extent of maser emission as a function of maser luminosity. The dashed line shows a power-law fit to the data with a power law exponent of 0.51.}\label{sizelumcorr}
\end{figure}

\begin{figure}
\centering
\includegraphics[width=0.45\textwidth]{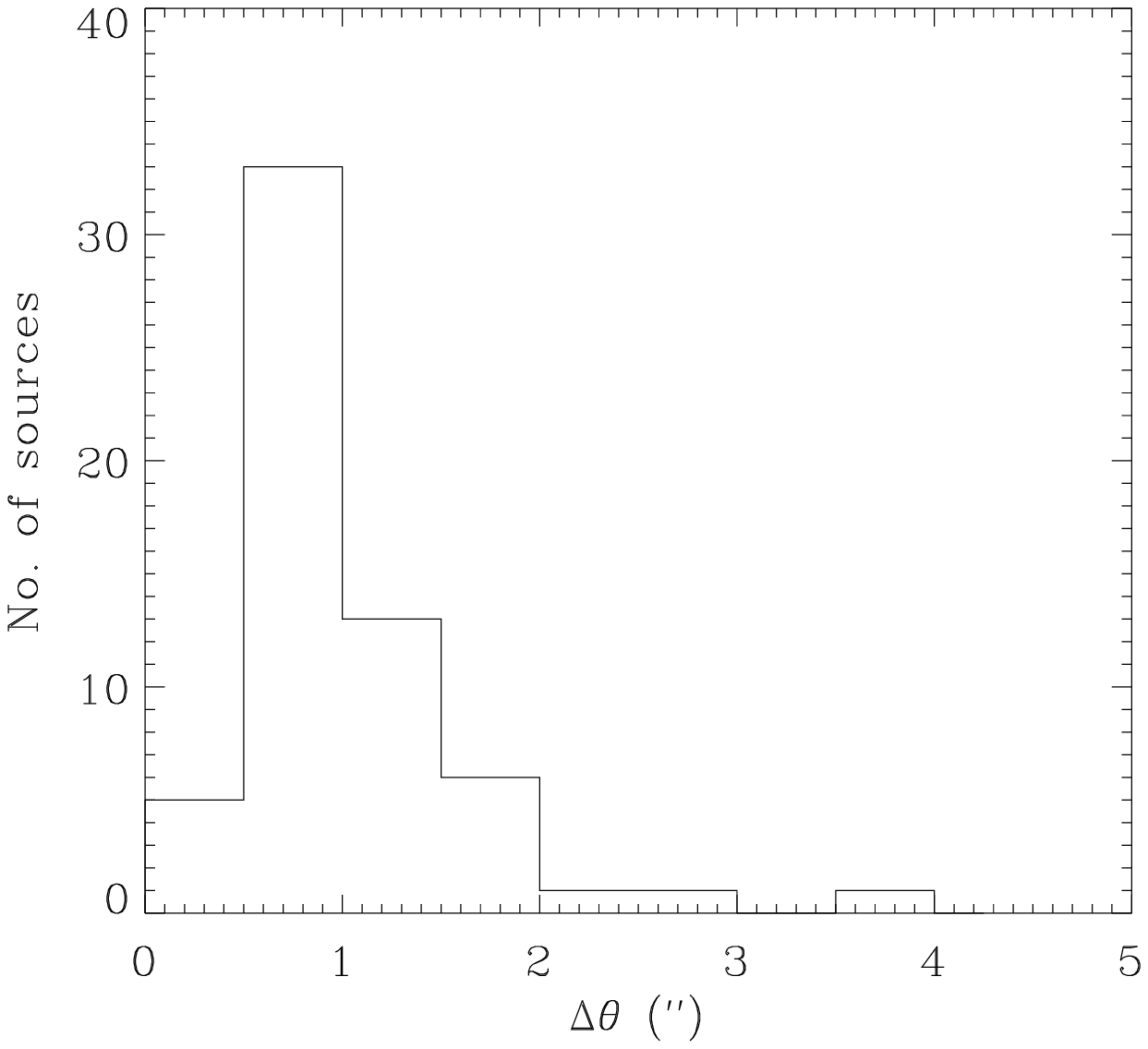}
\caption{Histogram of the separation between 6.7~GHz methanol masers and MIPS 24~$\mu$m point source counterparts.}\label{mipssephist}
\end{figure}

\end{document}